\documentclass[aps,letterpaper,showpacs,12pt]{revtex4}

\usepackage{amsmath}
\usepackage{bm}
\usepackage{graphicx}

\begin{document}

\preprint{\vbox{\hbox{JLAB-THY-03-193}}}

\title{ Modeling quark-hadron duality for relativistic, confined fermions}

\author{Sabine Jeschonnek$^{(1)}$ and J. W. Van Orden$^{(2,3)}$}

\affiliation{\small \sl (1) The Ohio State University, Physics
Department, Lima, OH 45804\\
(2) Jefferson Lab, 12000 Jefferson Ave, Newport News, VA 23606 \\
(3) Department of Physics, Old Dominion University, Norfolk, VA
23529}

\date{\today}

\begin{abstract}
We discuss a model for the study of quark-hadron duality in
inclusive electron scattering based on solving the Dirac equation
numerically for a scalar confining linear potential and a vector
color Coulomb potential. We qualitatively reproduce the features
of quark-hadron duality for all potentials considered, and discuss
similarities and differences to previous models that simplified
the situation by treating either the quarks or all particles as
scalars. We discuss the scaling results for PWIA and FSI, and the
approach to scaling using the analog of the Callan-Gross relation
for y-scaling.

\end{abstract}

\maketitle

\section{Introduction}

Quark-hadron duality was first discovered experimentally in
inclusive inelastic electron scattering by Bloom and Gilman
\cite{bgduality}, more than 30 years ago. In the past few years,
quark-hadron duality has generated a lot of interest both on the
experimental \cite{jlab,prdmom} and on the theoretical side
\cite{closeisgur,closezhao,closewallym,pp,
mpdirac,pps,donghe,dongli, kiev, leyaouancped, simula, ijmvo,
jvod2,elba}. Duality is a major point in the planned 12 GeV
upgrade of CEBAF at Jefferson Lab \cite{12gevwp}. It is also the
basis for using QCD sum rules \cite{qcdsr}, and plays an important
role in the study of semileptonic decays of heavy mesons
\cite{richural,semilep}.

Quark-hadron duality implies that in certain kinematic regions,
the appropriate average of hadronic observables is described by a
{\sl perturbative} quantum chromodynamics (pQCD) result. This is
of great practical interest, as we are actually able to carry out
a perturbative QCD calculation, in contrast to a full QCD or full
hadronic calculation. Surprisingly, duality was experimentally
shown to hold in inclusive inelastic electron scattering down to
momentum transfers of $Q^2 \approx 0.5 $ GeV$^2$ \cite{jlab}.
Duality also holds in the semi-leptonic decay of heavy quarks
\cite{isgurwise}, and in the annihilation $e^+e^- \rightarrow
hadrons$. The exact manner of averaging depends on the process.

Duality is not only a very interesting phenomenon by itself, but
it also has extremely important applications. As duality connects
the resonance region, i.e. the region where the final state
invariant mass $W < 2$ GeV, and the deep inelastic region, one may
infer information on one from the other. The earliest example
discussed was the extraction of the elastic nucleon form factor
from the deep inelastic scaling curve \cite{dgp}. In
\cite{jiunrau}, higher twist contributions were inferred from the
resonance data. This connection afforded by duality opens up the
large $x_{Bj}$ regime experimentally, as higher $Q^2$ measurements
are difficult to obtain - note that the $(e,e')$ cross section
contains the Mott cross section as a factor, and the Mott cross
section is proportional to $1/Q^4$. The large $x_{Bj}$ region is
much easier to access in the resonance region, as the necessary
$Q^2$ values there are much smaller.

One of the most exciting and promising applications of duality
will be the measurement of the neutron polarization asymmetry
$A_1^n$ at large $x_{Bj}$. There are many different theoretical
predictions for this quantity, ranging from 0 (unbroken SU(6)) to
1 (pQCD) \cite{nathana1n}. Experimental information on $A_1^n$ at
large $x_{Bj}$ would greatly enhance our understanding of the
valence quark spin distribution functions. There are very recent
new data from Jefferson Lab, going up to $x_{Bj} = 0.6$
\cite{a1data}, with small error bars, but the deep valence region
of $x_{Bj} \to 1$ remains inaccessible in deep inelastic
scattering.  If duality is well understood, one may take data in
the resonance region, apply a proper averaging procedure, and thus
obtain results for $A_1^n (x_{Bj}\rightarrow 1) $.

New theoretical approaches to a better understanding of duality
have been based on modeling: one branch uses the non-relativistic
constituent quark model, with some relativistic corrections, to
describe duality \cite{closeisgur,closewallym,donghe,dongli}, and
another branch starts the modeling with a relativistic one-body
equation \cite{closezhao,pp,mpdirac,pps,ijmvo,jvod2}.
The former branch makes contact with the phenomenology. It was
started by the pioneering work of Close and Isgur
\cite{closeisgur}, where the authors investigated how a summation
over the appropriate sets of nucleon resonances leads to parton
model results for the structure function ratios in the SU(6)
symmetric quark model. This work was recently expanded
\cite{closewallym} to include the effects of SU(6) spin-flavor
symmetry breaking. In \cite{donghe,dongli}, the authors considered
the first five low-lying resonances. Our results belong to the
latter branch. The goal of these modeling efforts is obvious: to
gain an understanding of quark-hadron duality and the conditions
under which it holds, by capturing just the essential physical
conditions of this rather complex phenomenon. We imposed these
basic requirements for a model: we require a  relativistic
description of confined valence quarks, and we treat the hadrons
in the infinitely narrow resonance approximation.

This paper is the third in a series on modeling quark-hadron
duality in inclusive electron scattering - the reaction in which
quark-hadron duality was first observed by Bloom and Gilman.
Previously, we simplified the situation by first assuming that all
particles involved are scalars \cite{ijmvo}, and then relaxed
these constraints for beam electrons and exchange photons, and
only assumed scalar quarks \cite{jvod2}. While these
simplifications are physically significant, they allowed us to
calculate all interesting quantities analytically or
semi-analytically. We found that the features of quark-hadron
duality were reproduced qualitatively in both models. Now, we have
taken one more step towards a realistic description of the
problem: previously, we simplified the problem by discussing
scalar quarks, but now, we use proper spin-1/2 quarks in our
model. This lays the foundation for investigating duality in
polarization observables.

Also, for the first time, in this paper we present calculations
for three different confining potentials. Previously, we used a
linear potential in the Klein-Gordon equation, which leads to a
"relativistic oscillator". Now, we present a scalar linear
confining potential and combine it with a vector potential: either
with a static color Coulomb potential or with a running color
Coulomb potential.

The paper is organized as follows: first, we introduce the model
and then give model results in PWIA.  The next section discusses
y-scaling: the connection between y-scaling and Bjorken scaling,
y-scaling results from our two previous models, y-scaling in PWIA
and FSI, and the sensitivity to different potentials. Then, we
discuss our results for sum rules in PWIA and FSI. In the next
section, we derive the analog of the Callan-Gross relation for
y-scaling, and investigate the onset of scaling through this
relation. Then, we summarize our results and give a brief outlook.

\section{The Model}

Our model consists of a constituent quark bound to an infinitely
heavy di-quark and is represented by the Dirac hamiltonian
\begin{equation}
\hat{H}=\bm{\alpha}\cdot\hat{\bm{p}}+\beta\left(m+V_s(r)\right)+V_v(r)\,,
\end{equation}
where the scalar potential is a linear confining potential given
by
\begin{equation}
V_s(r)=br,\qquad\qquad b=0.18{\rm GeV}^2\,.
\end{equation}
We have used the constituent quark mass in this paper, as our main
interest is the study of quark-hadron duality, which sets in at
rather low $Q^2$, experimentally $Q^2 \approx 0.5~GeV^2$ is
enough. In this kinematic region, the appropriate degree of
freedom is the constituent quark, which has acquired mass through
spontaneous chiral symmetry breaking. We have used a value for the
quark mass of $m = 258.46~MeV$ - obtained previously in a fit to
heavy mesons \cite{waw}. Calculations will be presented where the
vector color Coulomb potential is absent, that is $V_c(r)=0$,
where the vector potential is the simple static Coulomb potential
\begin{equation}
V_c(r)=-\frac{4}{3}\frac{\alpha_s}{r}
\end{equation}
with $\alpha_s = 0.181$ and where the color Coulomb potential is
corrected to allow for the running coupling constant in a manner
similar to that used by Godfrey and Isgur \cite{godfreyisgur}.
This potential has the form
\begin{equation}
V_{cr}(r)=-\frac{4}{3r}\left(
\alpha_c\frac{1+e^-\frac{\rho_0}{\delta}}{1+e^\frac{\sqrt{b}r-\rho_0}{\delta}}
+\sum_{i=1}^2\alpha_i {\rm erf}(\gamma_i r)\right)
\end{equation}
where
\begin{eqnarray}
\alpha_c &=& 0.118 \nonumber\\
\rho_0&=& 0.04\nonumber\\
\delta&=& 0.01 \nonumber\\
\alpha_1&=& 0.239\nonumber\\
\alpha_2&=& 0.271 \nonumber\\
\gamma_1 &=& 0.746\ {\rm GeV}\nonumber\\
\gamma_2 &=&  5.40\ {\rm GeV}
\end{eqnarray}

We assume that only the light quark carries a charge, and we
choose unit charge for the light quark for simplicity.  The
inclusive cross section is given by the usual Rosenbluth equation
\begin{equation}
\frac{d^2\sigma}{d\Omega dE}=\sigma_{\rm
Mott}\left\{\frac{Q^4}{\bm{q}^4}R_L(q,\nu)+
\left(\frac{Q^2}{2\bm{q}^2}+\tan^2\frac{\theta}{2}\right)
R_T(q,\nu)\right\}\,,
\end{equation}
where $\sigma_{\rm Mott}$ is the Mott cross section, $\bm{q}$ is
the three-momentum transfer from the electron to the target, $\nu$
is the energy transfer and $Q^2=\bm{q}^2-\nu^2$. The longitudinal
and transverse response functions for the model are given by
\begin{equation}
R_L(q,\nu)=\sum_f\left|\left<\Psi_f\right|e^{i\bm{q}\cdot\bm{r}}\left|
\Psi_0\right>\right|^2\delta\left(\nu+E_0-E_f\right)
\end{equation}
and
\begin{equation}
R_T(q,\nu)=\sum_f\sum_{i=1}^2\left|\left<\Psi_f\right|
e^{i\bm{q}\cdot\bm{r}}\alpha_i\left|
\Psi_0\right>\right|^2\delta\left(\nu+E_0-E_f\right)\,,
\end{equation}
where $E_0$ is the ground state energy. In terms of these response
functions the structure functions are
\begin{equation}
W_1(Q^2,\nu)=\frac{1}{2}R_T(q,\nu)
\end{equation}
and
\begin{eqnarray}
W_2(Q^2,\nu)=\frac{Q^4}{\bm{q}^4}R_L(q,\nu)
+\frac{Q^2}{2\bm{q}^2}R_T(q,\nu)\,.
\end{eqnarray}

The Dirac wave functions and energy eigenvalues are obtained by
integrating the Dirac equation using the Runge-Kutta-Feldberg
technique and solutions are obtained for energies up to 12 GeV
with the radial quantum number of $n\cong 200$ and $|\kappa|\leq
70$.

In our model, we excite the bound quark from the ground state to
higher energy states, and do not allow it to decay. Thus, we do
not include any particle production in our model, and are strictly
quantum-mechanical in this sense. We do not have any gluons in our
model, either, which means that we do not encounter any radiative
corrections. Since the response functions consist of a sum of
delta functions, we choose to smear out the response functions by
folding with a narrow gaussian for purposes of visualization. The
smeared response functions are then given by
\begin{equation}
R_{L(T)}(q,\nu)=\frac{1}{\sqrt{\pi}\epsilon}
\int_{-\infty}^{\infty}d\nu'\,e^{-\frac{(\nu-\nu')^2}{\epsilon^2}}
R_{L(T)}^{unsmeared}(q,\nu')\,. \label{smear}
\end{equation}

Before presenting numerical results, we would like to remind our
readers that, while the present model is more realistic than its
predecessors, its results should not be compared {\sl
quantitatively} to inclusive electron scattering from a nucleon.
Due to the assumption of an infinitely heavy antiquark (or
diquark) to which the light quark is bound, our calculation most
resembles inclusive electron scattering from a B-meson, which has
never been measured. However, the goal of our work is to gain a
qualitative understanding of duality, and the current
simplification is no impediment to this.

\section{The Plane Wave Impulse Approximation}

The analog to pQCD for this model is the plane wave impulse
approximation where the bound quark is knocked into the continuum
by the absorption of the virtual photon. The response tensor for
this approximation is
\begin{eqnarray}
W^{\mu\nu}&=&\frac{1}{8}\int\frac{d^3p}{(2\pi)^3}
\frac{m}{E_{\bm{p}+\bm{q}}}\delta(\nu+E_0-E_{\bm{p}+\bm{q}})\nonumber\\
&&\times {\rm Tr}\left[\gamma^\mu
\frac{\gamma\cdot(p+q)+m}{2m}\,\gamma^\nu \left(\gamma\cdot
n_v(p)+n_s(p)\right)\right]
\end{eqnarray}
where the vector and scalar momentum density distributions
$n_v(p)$ and $n_s(p)$ are defined in terms of the ground state
wave function
\begin{equation}
\Psi_{10\frac{1}{2}m}(\bm{p})=\left(\begin{array}{c}
\psi^{(+)}_{10\frac{1}{2}}(p){\cal Y}_{0\frac{1}{2}}^m(\Omega_p)\\
\psi^{(-)}_{10\frac{1}{2}}(p){\cal Y}_{1\frac{1}{2}}^m(\Omega_p)
\end{array}\right)
\end{equation}
as
\begin{equation}
n_v(p)=\left(n_v^0(p), \frac{\bm{p}}{|\bm{p}|}n_v^s(p)\right)
\end{equation}
with
\begin{equation}
n_v^0(p)=\frac{1}{2\pi}\left({\psi^{(+)}_{10\frac{1}{2}}}^2(p)+
{\psi^{(-)}_{10\frac{1}{2}}}^2(p)\right)
\end{equation}
and
\begin{equation}
n_v^s(p)=\frac{1}{\pi}\psi^{(+)}_{10\frac{1}{2}}(p)
\psi^{(-)}_{10\frac{1}{2}}(p)\,;
\end{equation}
and
\begin{equation}
n_s(p)=\frac{1}{2\pi}\left({\psi^{(+)}_{10\frac{1}{2}}}^2(p)-
{\psi^{(-)}_{10\frac{1}{2}}}^2(p)\right)\,.
\end{equation}

After performing the angular integrals, the response functions can
be written as
\begin{eqnarray}
R_L(q,\nu)&=&\frac{1}{16\pi^2q}\int^{y+2q}_{|y|}dp\,
p\biggl\{(\nu+E_0)n_v^0(p)+mn_s(p)\nonumber\\
&&+\left.
\frac{(\nu+E_0)^2+p^2-q^2-m^2}{2p}n_v^s(p)\right\}\label{PWIAL}
\end{eqnarray}
and
\begin{eqnarray}
R_T(q,\nu)&=&\frac{1}{8\pi^2q}\int^{y+2q}_{|y|}dp\,
p\biggl\{(\nu+E_0)n_v^0(p)-mn_s(p)\nonumber\\
&&-\left. \frac{(\nu+E_0)^2-p^2+q^2-m^2}{2q}\
\frac{(\nu+E_0)^2-p^2-q^2-m^2}{2pq}n_v^s(p)\right\}\,,
\label{PWIAT}
\end{eqnarray}
where
\begin{equation}
y=\sqrt{(\nu+E_0)^2-m^2}-q\,.\label{defy}
\end{equation}

\section{y Scaling}

Duality implies that we see scaling at high enough energy and
momentum transfers, and that the results in the resonance region,
for lower momentum transfers, oscillate around this scaling curve
\cite{jlab}. So, the first point that needs to be checked is the
onset of scaling in our model. For our previous two models, we
were able to show analytically that there is scaling, and that the
scaling curves for PWIA and FSI coincide exactly. This result is
in contrast to \cite{pp}, where a $30 \%$ difference was found
between the PWIA and FSI scaling curves. In \cite{pp}, a different
one-body equation, a semi-relativistic Hamiltonian of the form $H
= \sqrt{|\bm{p}|^2} + \sqrt{\sigma} r$ for massless quarks was
used, whereas we have used a Klein-Gordon equation in our previous
modeling, appropriate for spinless ``quarks" with mass. Now that
we use the Dirac equation, we will again have to investigate the
important question if the two scaling curves coincide or not. The
scaling behavior, and a possible violation of ``FSI scaling curve
= PWIA scaling curve" would have implications for our
interpretation of deep inelastic scattering data, where one
usually assumes that FSI can be neglected. The question if and how
scaling may arise in the presence of FSI has been discussed in the
literature before, both for non-relativistic
\cite{greenberg,gurvitzrinat,harrington} and relativistic
\cite{lps} approaches. Here, we are interested in duality, and
need to compare the FSI scaling results with the PWIA scaling
results.

Note that there is a certain arbitrariness in defining ``scaling"
- have we reached scaling only once all curves coincide perfectly,
or is a minimal shift in the curves when, e.g., doubling the
momentum, enough? In the following, our usage is that ``scaling"
is reached when the change in the curve is minimal for a very
substantial change in the momentum transfer.

\subsection{Bjorken scaling and y-scaling}

\begin{figure}[h]
\includegraphics[width=20pc,angle=270]{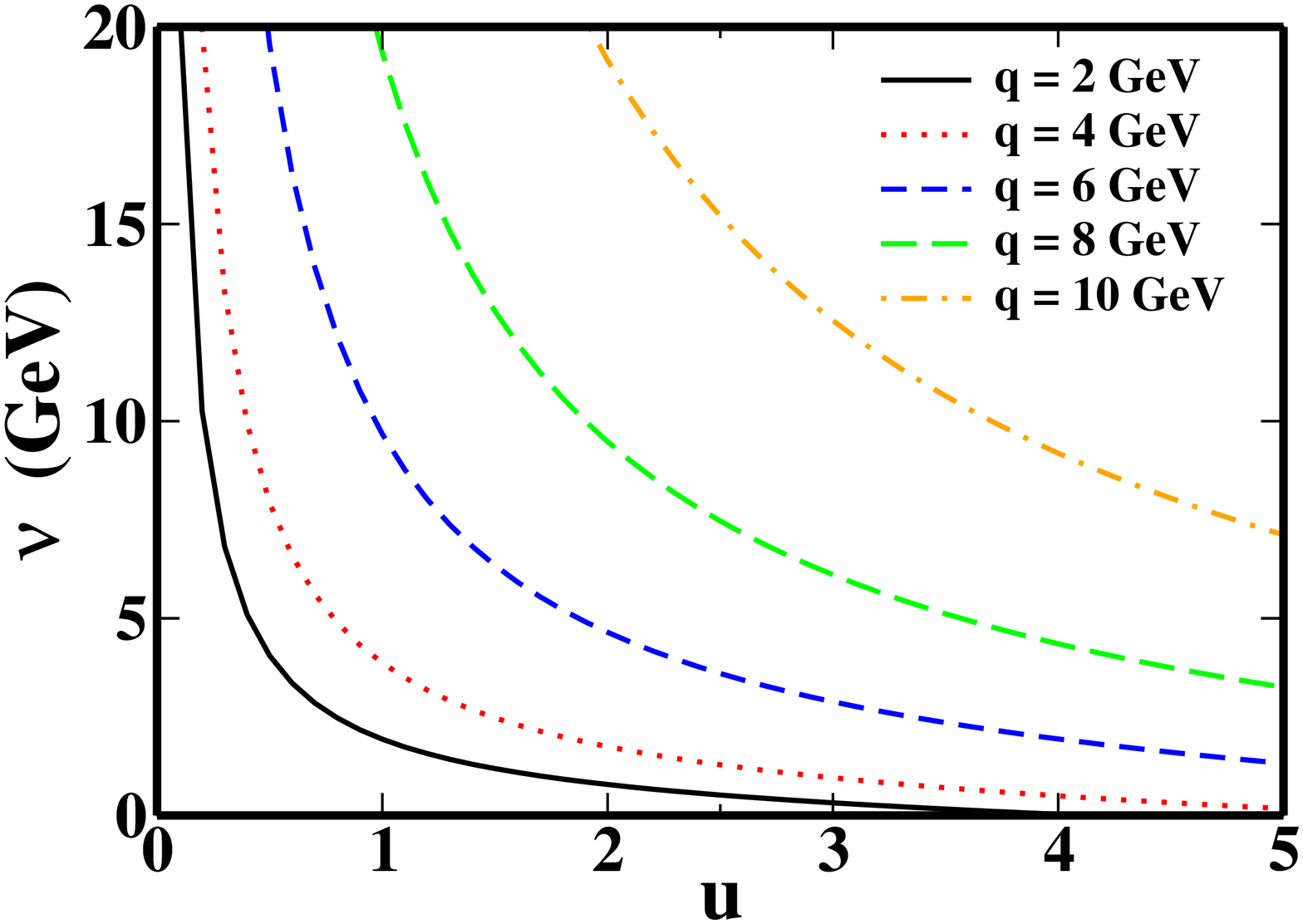}
\includegraphics[width=20pc,angle=270]{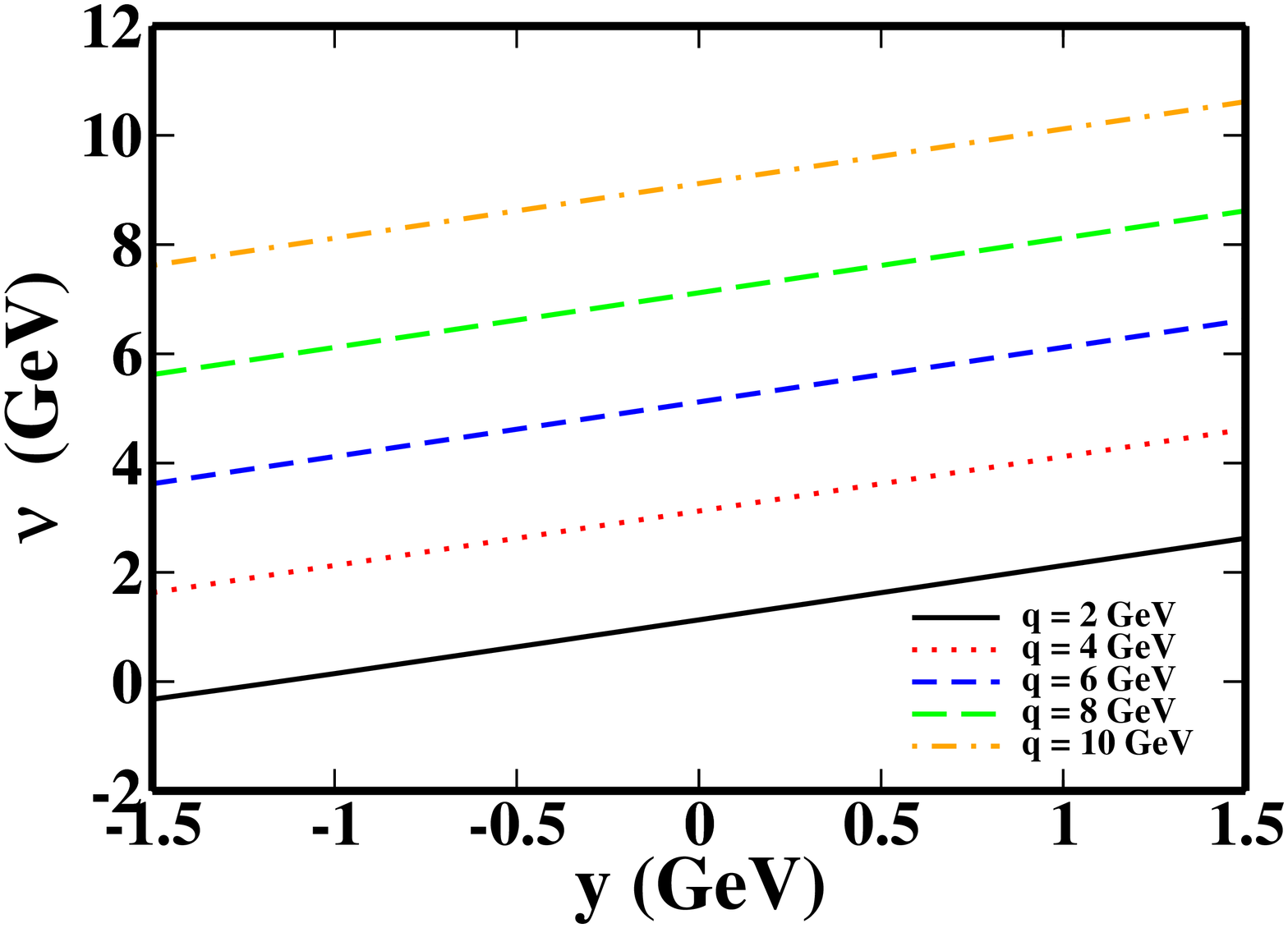}
\caption{The transferred energy $\nu$ as a function of the scaling
variable $u$ for various values of the four-momentum transfer
$Q^2$ (top panel), and as a function of the scaling variable $y$
for various values of the three-momentum transfer $q$ (bottom
panel).} \label{fignuofuy}
\end{figure}

In our two previous papers on modelling duality
\cite{ijmvo,jvod2}, we have presented our results for the scaling
functions at fixed $Q^2$ as functions of a scaling variable $u$.
In the Bjorken limit, $Q^2 \to \infty$, $u$ goes to $u_{Bj} =
\frac{M}{m} x_{Bj}$, where $x_{Bj} = \frac{Q^2}{2 M \nu}$ is
Bjorken's scaling variable, and $u_{Bj}$ is the appropriately
rescaled version of $x_{Bj}$ for the case of a a target with
infinite mass $M$. The values of the energy transfer $\nu$
accessed for various, fixed values of $Q^2$ are shown in
Fig.~\ref{fignuofuy} in the left panel.

We would like to point out that the previously used simplification
of modelling the quark as a scalar has allowed us to obtain
analytic results for the structure functions, giving us access to
high energy transfers and high $Q^2$ without any practical,
numerical problems. Now, however, we have to rely on a solely
numerical solution of the Dirac equation, and cannot push to
arbitrarily high $Q^2$ and $\nu$ values anymore. In particular, we
find about 28000 energy eigenstates, all below $12~GeV$ energy.
While this is an impressive number of states, it is clear from
Fig.~\ref{fignuofuy} that we will not be able to reach the high
$Q^2$ values found necessary for scaling in our model \cite{jvod2}
with energy transfers of less than $12~GeV$. Note that the peak of
the structure function $ \nu W_2$ is localized around $u \approx 2
- 3$, so the higher $Q^2$ values accessible at larger $u$ have
little practical significance.
\begin{figure}[ht]
\includegraphics[width=20pc,angle=270]{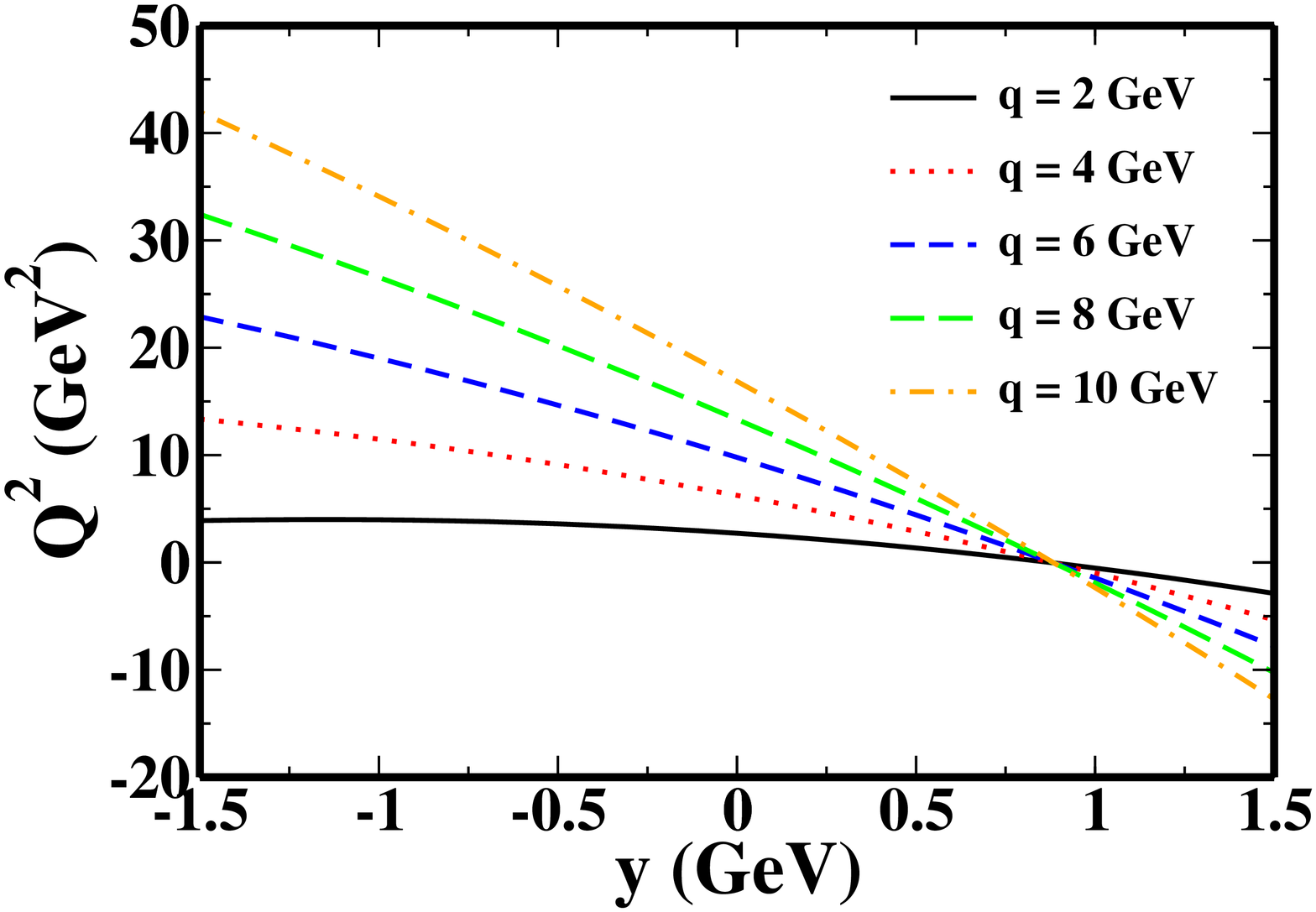}
\caption{The transferred four-momentum $Q^2$ as a function of the
scaling variable $y$ for various values of the three-momentum
transfer $q$.} \label{figqsqofy}
\end{figure}

The matrix elements take their simplest form if calculated as
functions of $q$. Therefore, we consider y-scaling in this paper,
and use the scaling variable as defined in Eq.(\ref{defy}),

$$
y = \sqrt{(\nu + E_0)^2 - m^2} - q \, . $$
This variable leads to scaling for fixed $|\bf{q}| $  $= q $
\cite{elba}. The kinematics accessed in a y-scaling analysis are
shown in Figs.~\ref{fignuofuy} and \ref{figqsqofy}.
Fig.~\ref{fignuofuy}, bottom panel, shows the values of the energy
transfer $\nu$ accessed in the range from $y = - 2~GeV$ to $y = 2~
GeV$ - this is the range in which we have non-negligible
contributions to our response functions and structure functions.
Fig.~\ref{figqsqofy} shows that we are now probing a different
kinematic region than in $u$-scaling: we access both spacelike and
timelike values of the four-momentum transfer $Q^2$. At lower,
negative $y$ values, $Q^2$ is spacelike and rather large. In this
region, we can easily reach large values associated with scaling
in $u$ or $x$-type variables. The accessed $Q^2$ values decrease
slowly with increasing $y$, and at a point where $y_0 = E_0 +
O(\frac{1}{q})$, we reach the photopoint, $Q^2 = 0$. Note that due
to the fact that $y_0$ is independent of $q$ in first
approximation, the photopoint is reached at the same $y$ value
independent of the considered three-momentum transfer $q$. For $y
> y_0$, we probe the timelike region. Note that one also reaches
timelike values of $Q^2$ for $y < -2 q - E_0 + O(\frac{1}{q})$; in
this region, however, we find zero strength of the responses, and
it is practically irrelevant for our case.

\subsection{Comparison with previous results}

In order to provide a connection between $u$-scaling and
$y$-scaling, we show our results for our two previous models with
scalar quarks for $y$-scaling.
We start out with the all-scalar results discussed in
\cite{ijmvo}. In \cite{ijmvo}, we treated beam, exchange particle
and quark target as scalars. This simplified treatment allows for
an analytic solution of the problem. With all particles being
scalars, only one structure function appears, there is little
structure, and scaling sets in around $Q^2 \approx 15~GeV^2$.
Scaling for the PWIA sets in at very low $Q^2$ of about $2~GeV^2$.
Keep in mind that we scatter from an infinitely heavy target, so
that a comparison of numerical values for $Q^2$ for the onset of
scaling with actual nucleon data is meaningless.

\begin{figure}[ht]
\includegraphics[width=20pc,angle=270]{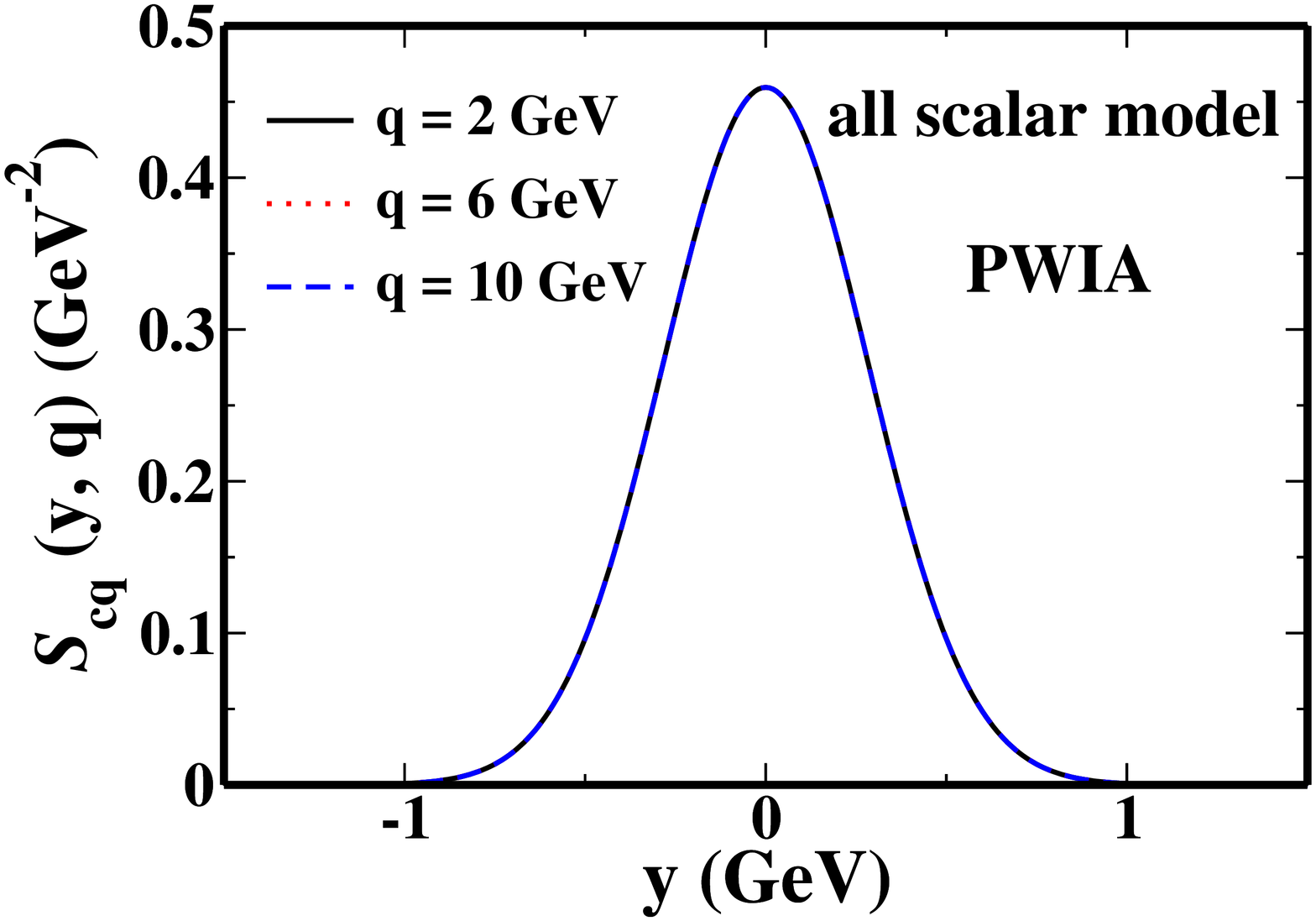}
\includegraphics[width=20pc,angle=270]{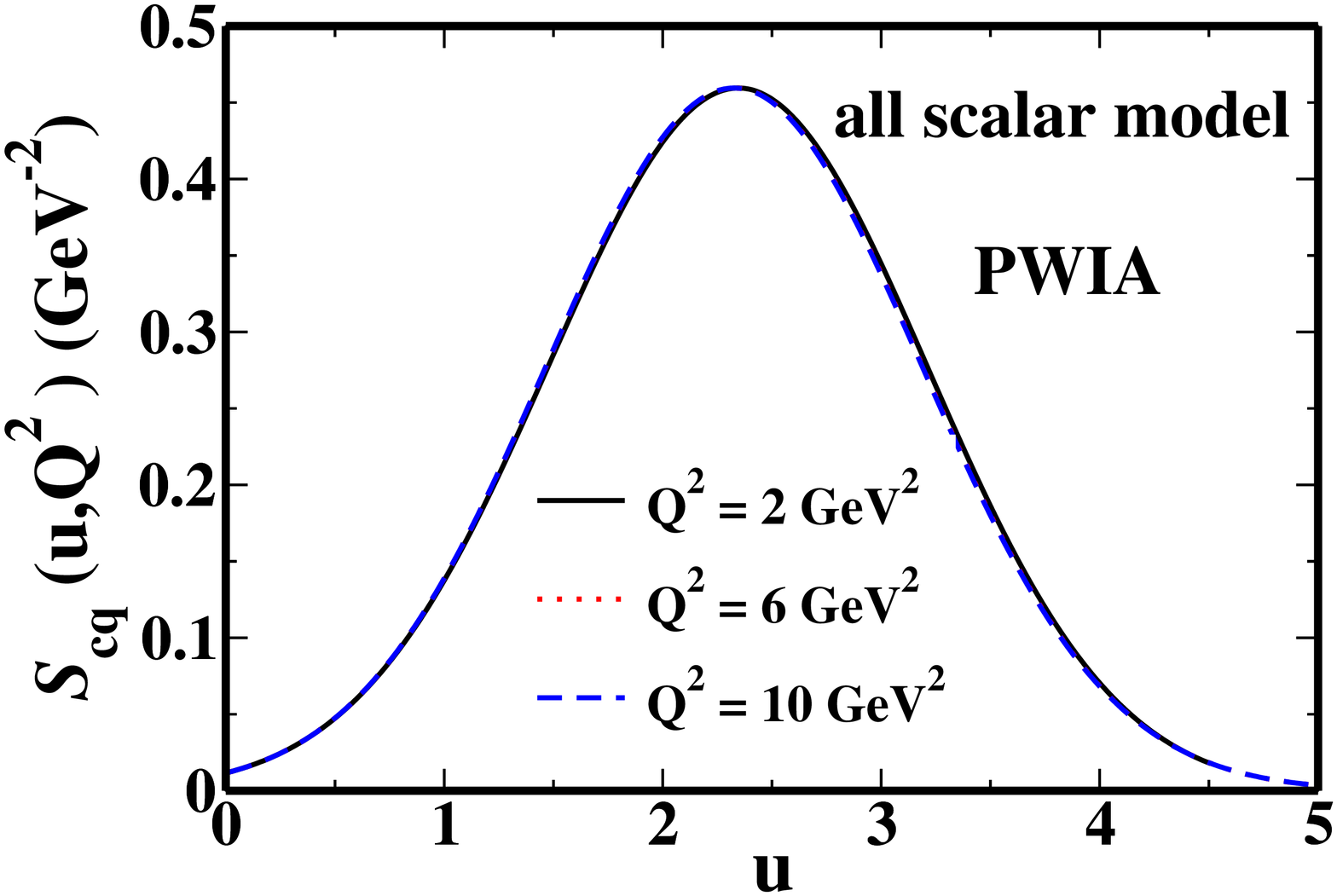}
\caption{The scaling function ${\cal{S}}_{cq}$ of the all-scalar
model {\protect{\cite{ijmvo}}} is plotted versus $y$ for several
values of the three-momentum transfer $q$ in the top panel, and it
is plotted versus $u$ for several values of the four-momentum
transfer $Q^2$ in the bottom panel. The results shown have been
calculated in PWIA.} \label{figaspwia}
\end{figure}
Fig.~\ref{figaspwia} (top panel) shows the corresponding
$y$-scaling plot for the PWIA. One can see clearly that for $q =
2~GeV$, scaling has already set in. Obviously, there are no
resonance bumps in the PWIA plot, as the final state is a
fictitious ``free quark". In the bottom panel of
Fig.~\ref{figaspwia}, we show the same PWIA scaling function,
plotted for fixed four-momentum transfer $Q^2$ versus $u$. The
overall features of the curves, plotted either way, are the same:
they are smooth and scale quickly. In the $u$-scaling plot, one
sees that scaling has set in at $Q^2 = 6~GeV^2$, and the changes
from the lower value of $Q^2 = 2~GeV^2$ to  $Q^2 = 6~GeV^2$ are
tiny.

\begin{figure}[ht]
\includegraphics[width=20pc,angle=270]{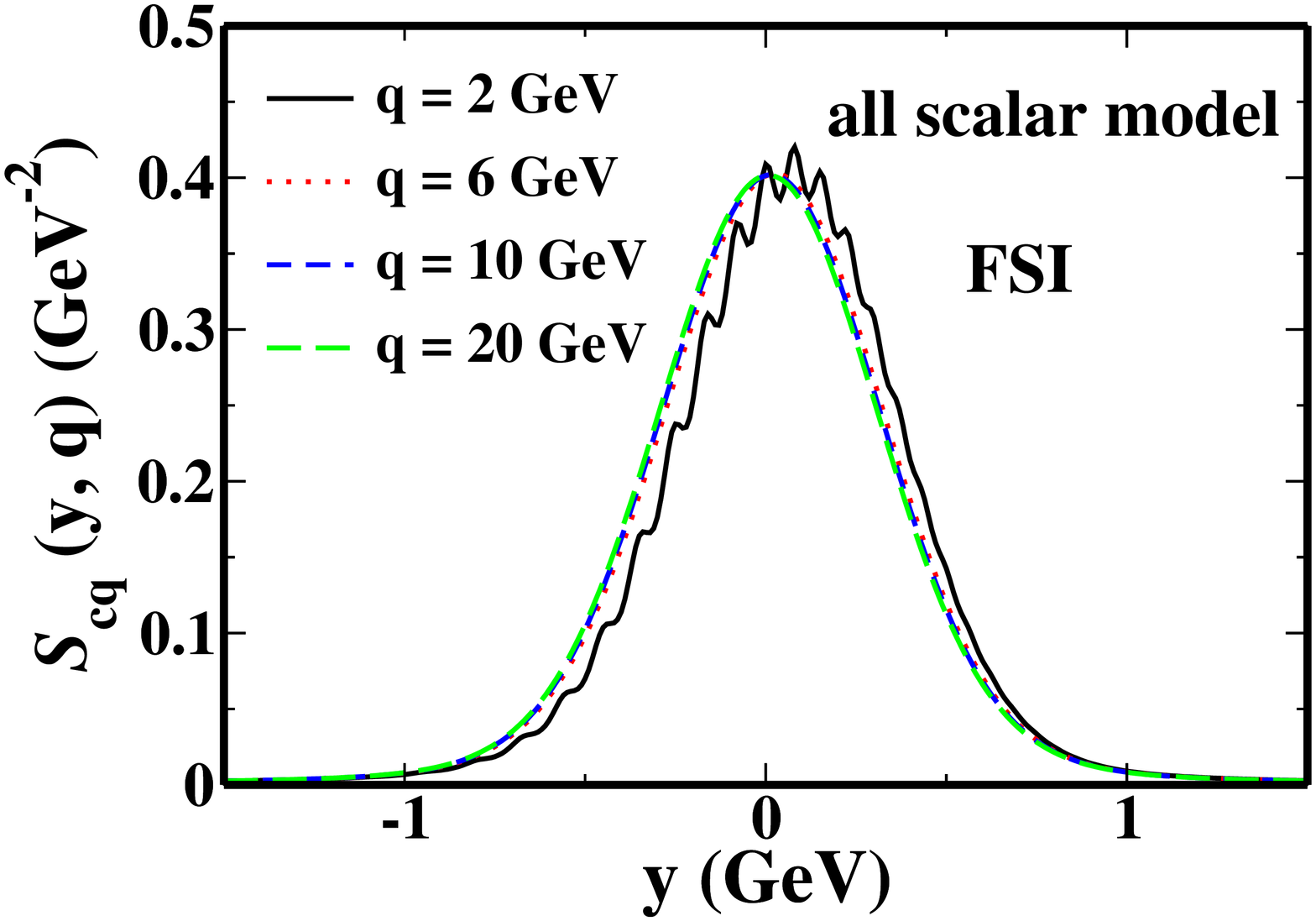}
\includegraphics[width=20pc,angle=270]{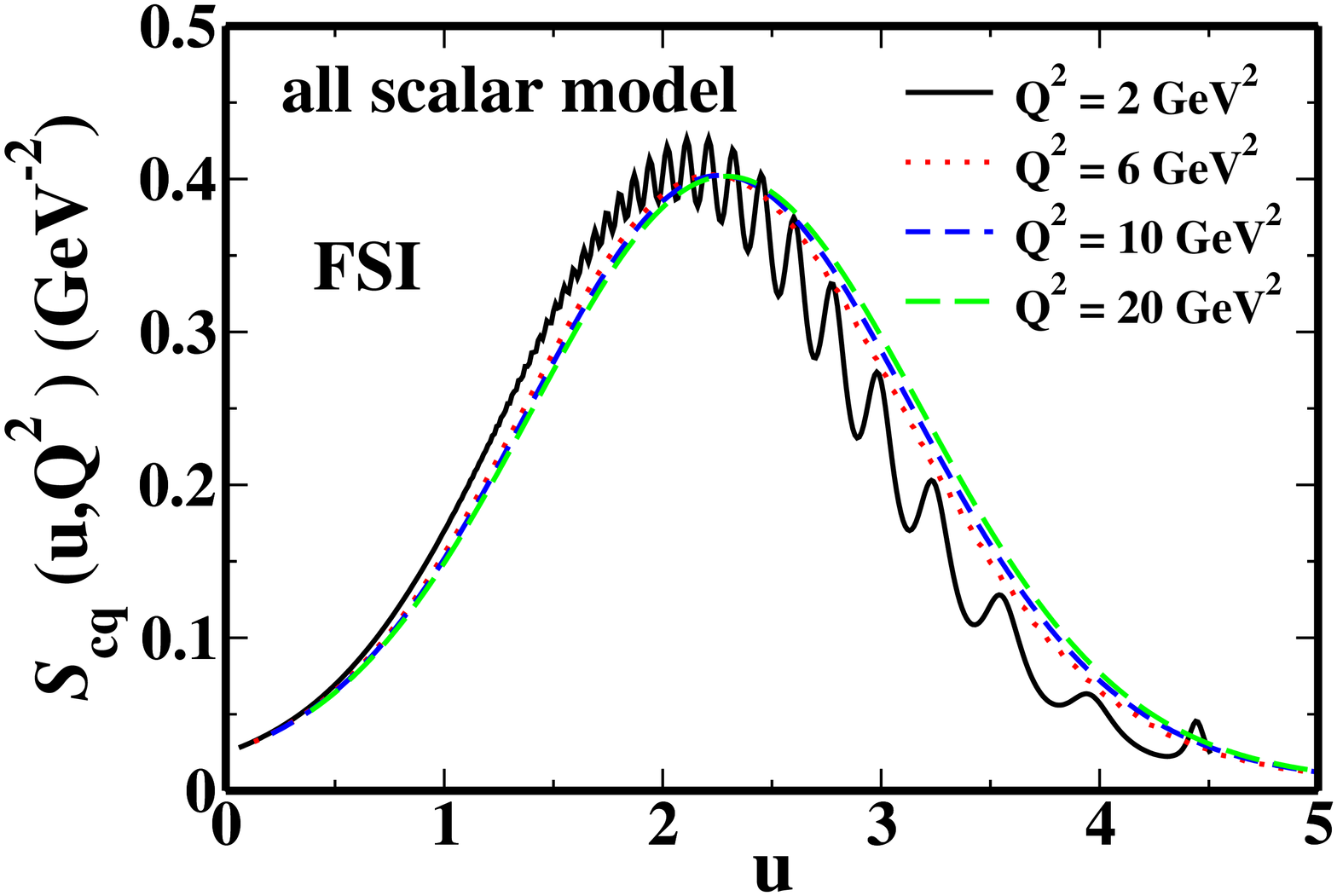}
\caption{The scaling function ${\cal{S}}_{cq}$ of the all-scalar
model is plotted versus $y$ for several values of the
three-momentum transfer $q$ in the top panel, and it is plotted
versus $u$ for several values of the four-momentum transfer $Q^2$
in the bottom panel. The results shown have been calculated
including final state interactions (FSI).} \label{figas}
\end{figure}
Fig.~\ref{figas}, top panel, shows the $y$-scaling plot for the
all-scalar model including FSI. Here, one can see that scaling
does take a while to set in - the lowest value for $q$ that is
displayed, $q = 2~GeV$, still shows the typical resonance bump
structure. However, the overall shape even at low $q$ is already
very close to the scaling curve, but slightly shifted towards
higher $y$ values. The scaling curve is approximated reasonably at
$q = 6~GeV$, and the curves start to coincide with each other once
$q = 10~GeV$ is reached.

This is contrasted with the $u$-scaling plot for the all-scalar
model including FSI in the bottom panel of Fig.~\ref{figas}. One
can see how the resonance bumps at lower $Q^2$ give way to smooth
curves at larger values of the four-momentum transfer. One can see
that scaling sets in just in the same way, independent of which
set of kinematic variables one chooses for plotting. For practical
reasons - the current matrix elements take their simplest form as
a function of $q$ - we choose to present our results as functions
of $q$ and $y$.

\begin{figure}[htp]
\includegraphics[width=20pc,angle=270]{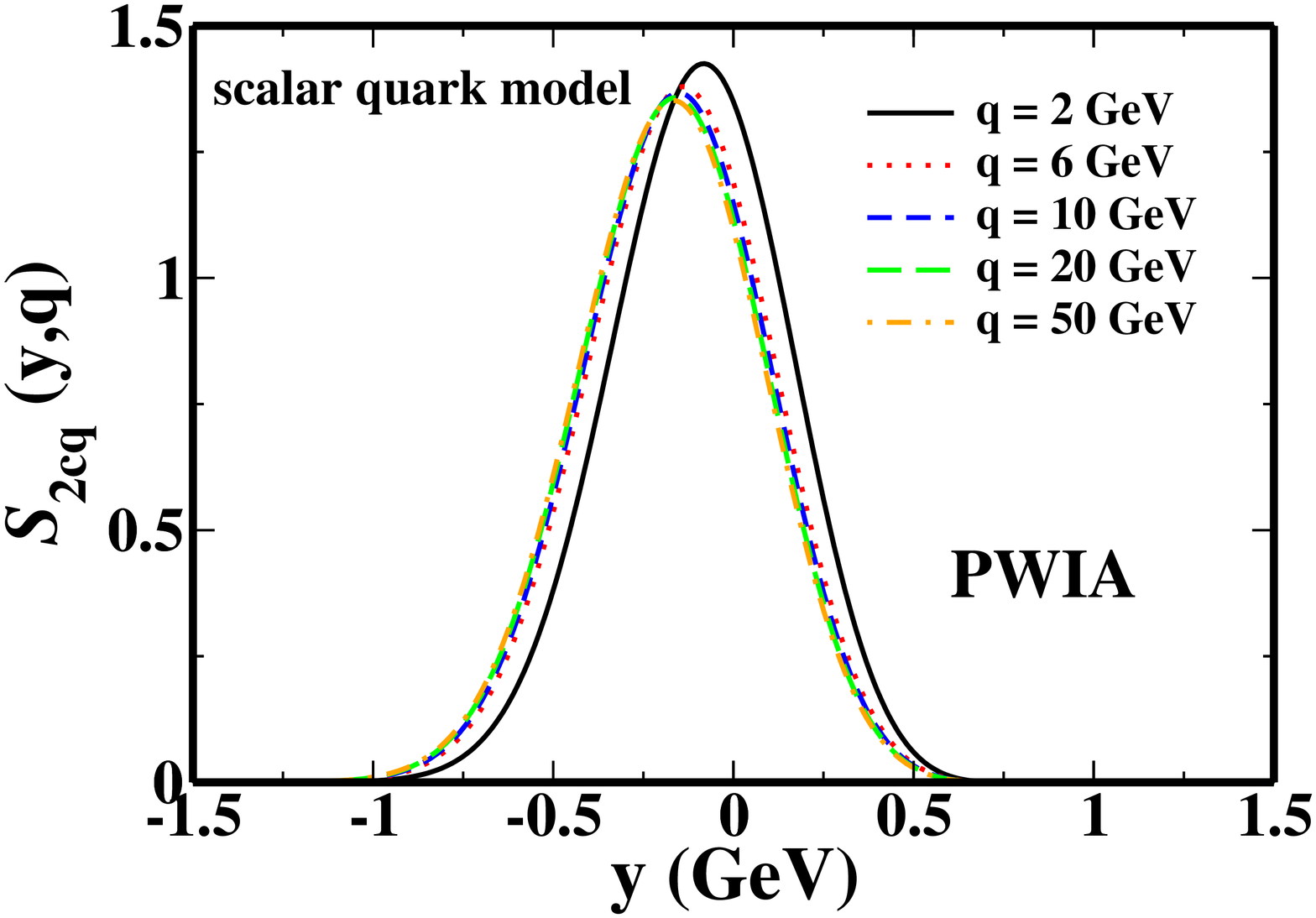}
\includegraphics[width=20pc,angle=270]{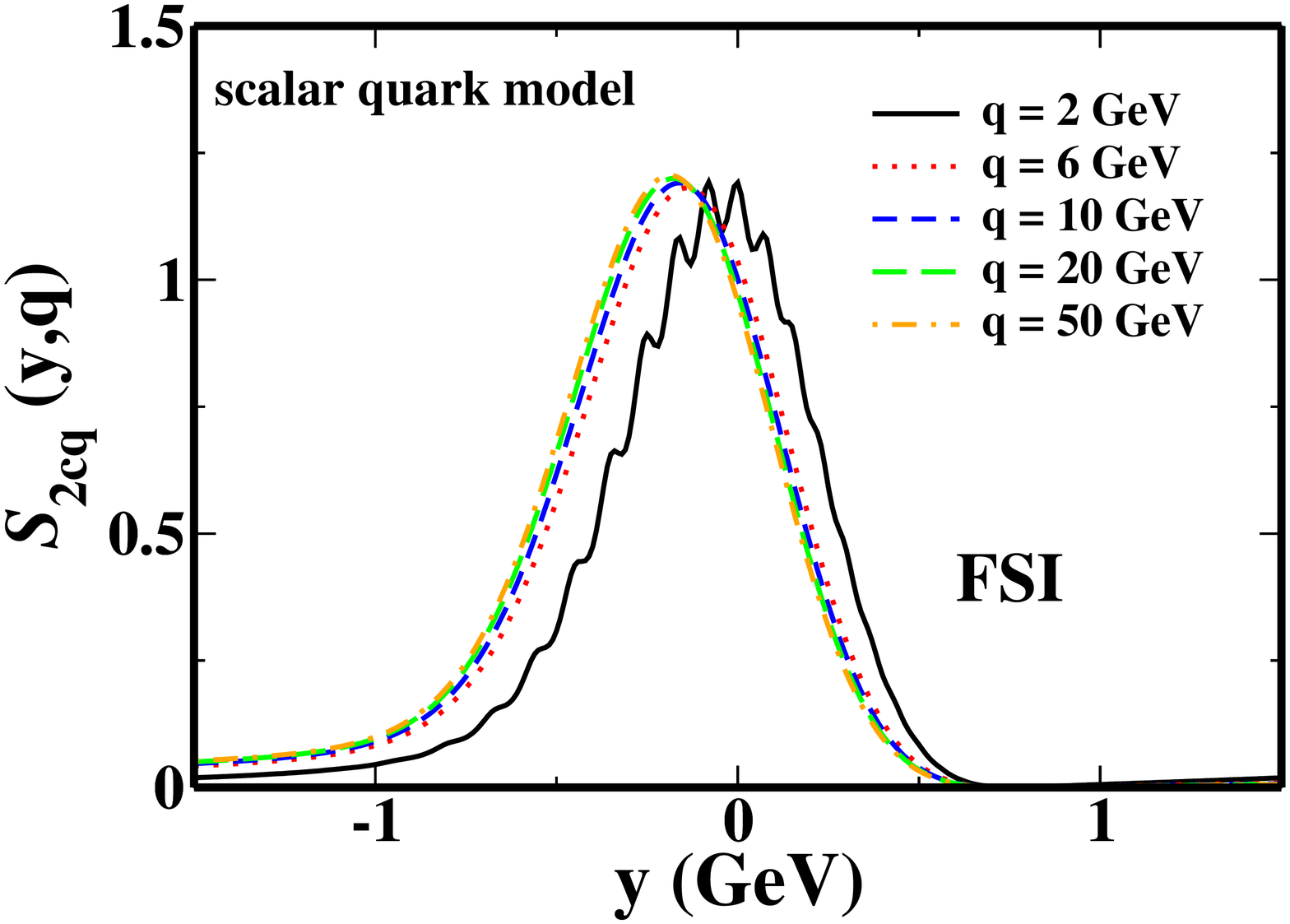}
\caption{The scaling function ${\cal{S}}_{2cq}$ of the ``scalar
quark" model is plotted versus $y$ for several values of the
three-momentum transfer $q$. The results shown in the top panel
have been calculated in PWIA, i.e. for the bound-free transition.
The results shown in the bottom panel have been calculated
including final state interactions (FSI), i.e. for the bound-bound
transition.} \label{figccy}
\end{figure}

Now, we will proceed to show the $y$ scaling results for the
"scalar quark model" \cite{jvod2}, too. This will allow us to
study the transition from a simple model to a more sophisticated
version, and to point out differences and common features. In
\cite{jvod2}, we included the proper spins for the beam electrons
and the exchange photons, but still treated the quark as a scalar.
We will refer to this model as the "scalar quark model" in the
present paper. This model has a conserved current, and a much
richer structure, due to the fact that the photon can have
transverse and longitudinal polarization, which leads to two
structure functions. In Fig.~\ref{figccy}, top panel, the PWIA
results definitely scale more slowly than for the all scalar
model. Scaling is reached at $q = 20~GeV$. Not surprisingly, the
FSI results, shown in the bottom panel of Fig.~\ref{figccy},
exhibit an even slower scaling. Even from $q = 20~GeV$ to $q =
50~GeV$, a small difference is visible in the scaling curves. This
is the slowest onset of $y$-scaling which we have encountered so
far, and this mirrors precisely what we expected from our previous
$u$-scaling results. Now we are in a good position to progress to
the main topic of this paper, the results for proper spin-$1/2$
quarks. Our study of the onset of scaling for PWIA and FSI will
allow us to draw some conclusions about the onset of $u$ scaling
for the FSI case, even if we can't calculate for the necessary
$Q^2$ numerically. First, we are going to present results for the
PWIA for three different potentials.

\subsection{y-scaling in PWIA}
\label{secpwiayscal}

First, we show our results for PWIA, where we expect scaling to
set in at lower momentum transfers than for the FSI results. For
the FSI results, the scaling curve is generated in our model by
many overlapping resonances, and one needs to reach kinematics
where a sufficient number of resonances is accessible to see
scaling.

The scaling variable y is defined by (\ref{defy}) and by inverting
to find $\nu$ as a function of $y$ and $q$, the response functions
can be written as functions of $q$ and $y$.  For example the PWIA
response functions (\ref{PWIAL}) and (\ref{PWIAT}) can be
rewritten as
\begin{eqnarray}
R_L(q,y)&=&\frac{1}{16\pi^2q}\int_{|y|}^{y+2q}dp p
\biggl\{\sqrt{(y+q)^2+m^2}n^0_v(p)+m\,n_s(p)\nonumber\\
&&\left.+ \frac{y^2+2qy+p^2}{2p}\,n^s_v(p)\right\}
\end{eqnarray}
and
\begin{eqnarray}
R_T(q,y)&=&\frac{1}{8\pi^2q}\int_{|y|}^{y+2q}dp p
\biggl\{\sqrt{(y+q)^2+m^2}n^0_v(p)-m\,n_s(p)\nonumber\\
&&\left.- \frac{y^2+2qy+2q^2-p^2}{2q}
\,\frac{y^2+2qy-p^2}{2pq}\,n^s_v(p)\right\}\,.
\end{eqnarray}

In the limit of large $q$ the PWIA response functions become
\begin{eqnarray}
\lim_{q\rightarrow\infty}R_L(q,y)&=&\frac{1}{16\pi^2}\int_{|y|}^{\infty}dp
p \left\{n^0_v(p)+ \frac{y}{p}\,n^s_v(p)\right\}
\end{eqnarray}
and
\begin{eqnarray}
\lim_{q\rightarrow\infty}R_T(q,y)&=&\frac{1}{8\pi^2}\int_{|y|}^{\infty}dp
p \left\{n^0_v(p)- \frac{y}{p}\,n^s_v(p)\right\}\,.
\end{eqnarray}
These response functions therefore scale in $y$. Note that the
longitudinal response is shifted toward positive $y$ while the
transverse is shifted toward negative $y$, since $n_v^s(p)$ is
positive. The overall peak heights of the longitudinal and
transverse responses in PWIA have roughly a $1:2$ ratio.

\begin{figure}
\includegraphics[width=20pc,angle=270]{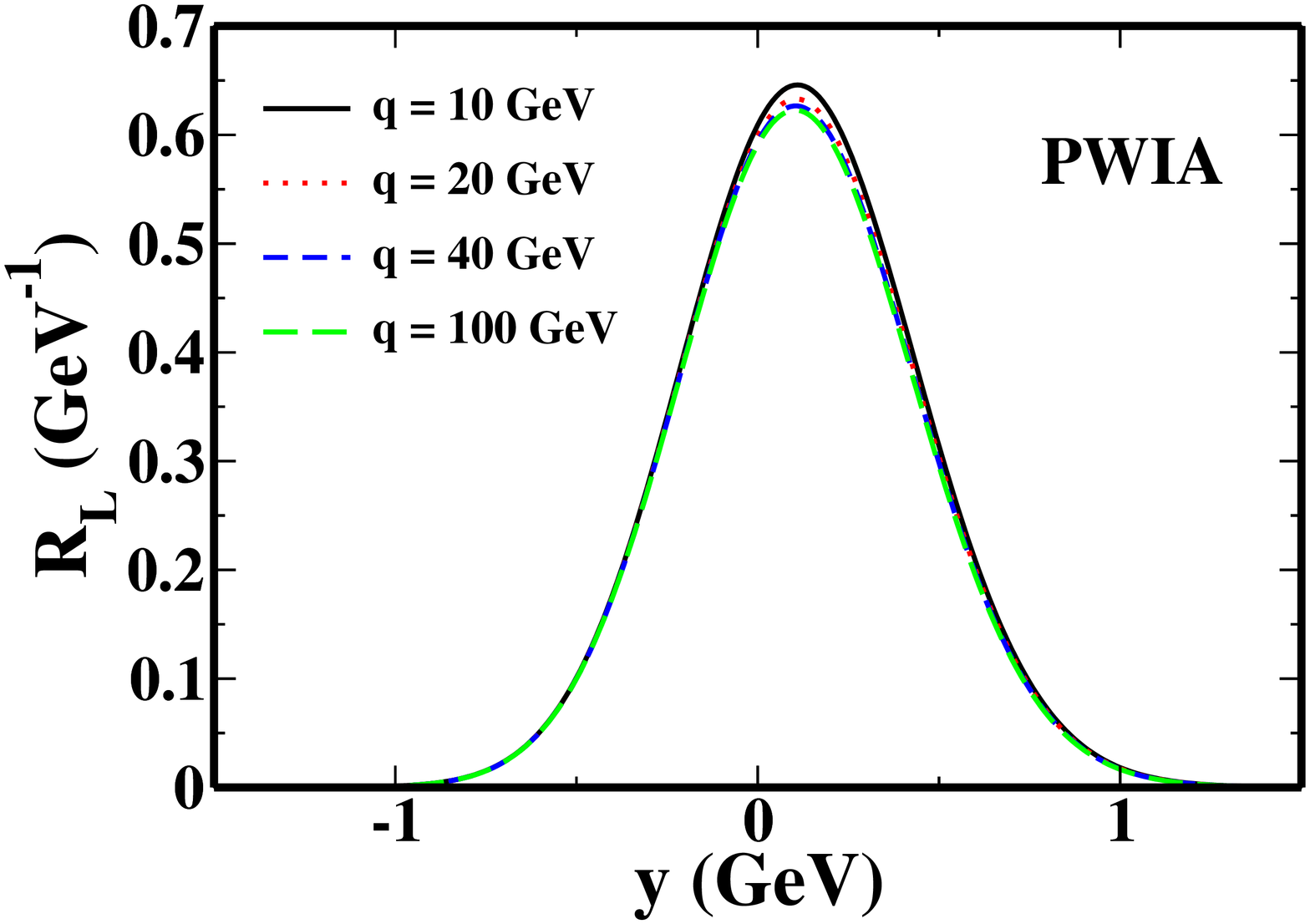}
\includegraphics[width=20pc,angle=270]{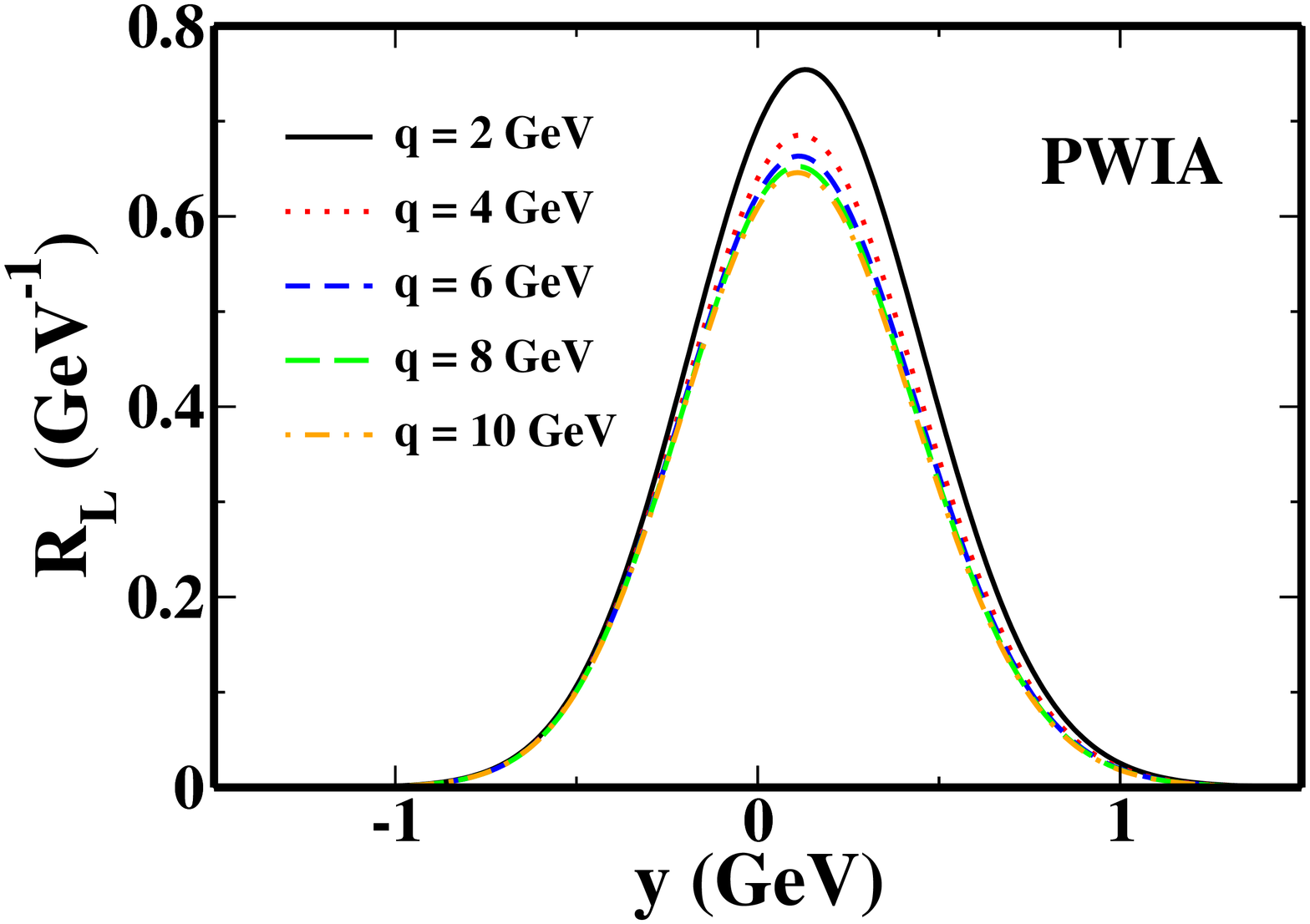}
\caption{The longitudinal response function $R_L$ is plotted
versus $y$ for several low (bottom panel) and high (top panel)
values of the three-momentum transfer $q$. The results shown have
been calculated in PWIA, using the linear potential.}
\label{figpwlinrl}
\end{figure}
\begin{figure}
\includegraphics[width=20pc,angle=270]{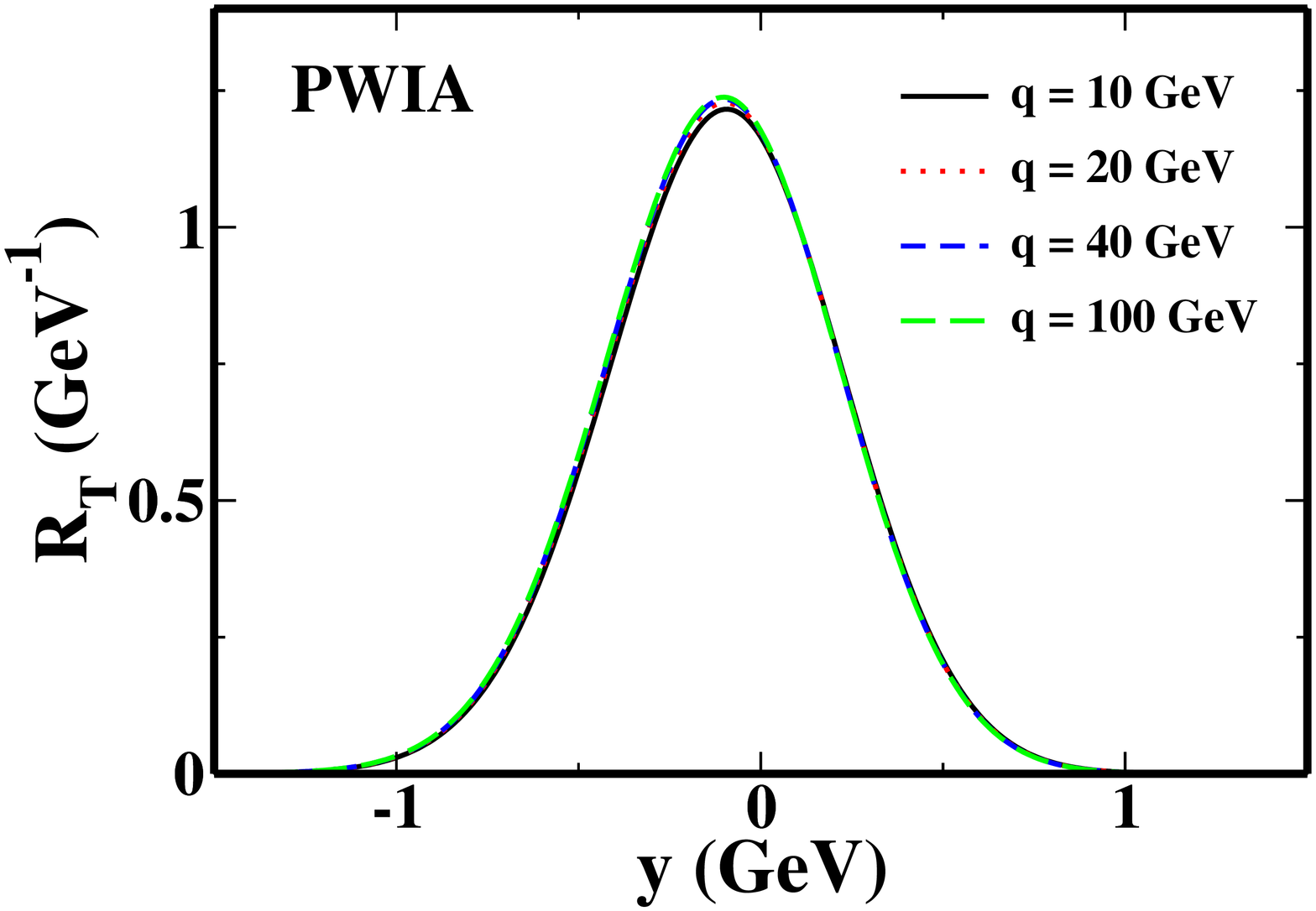}
\includegraphics[width=20pc,angle=270]{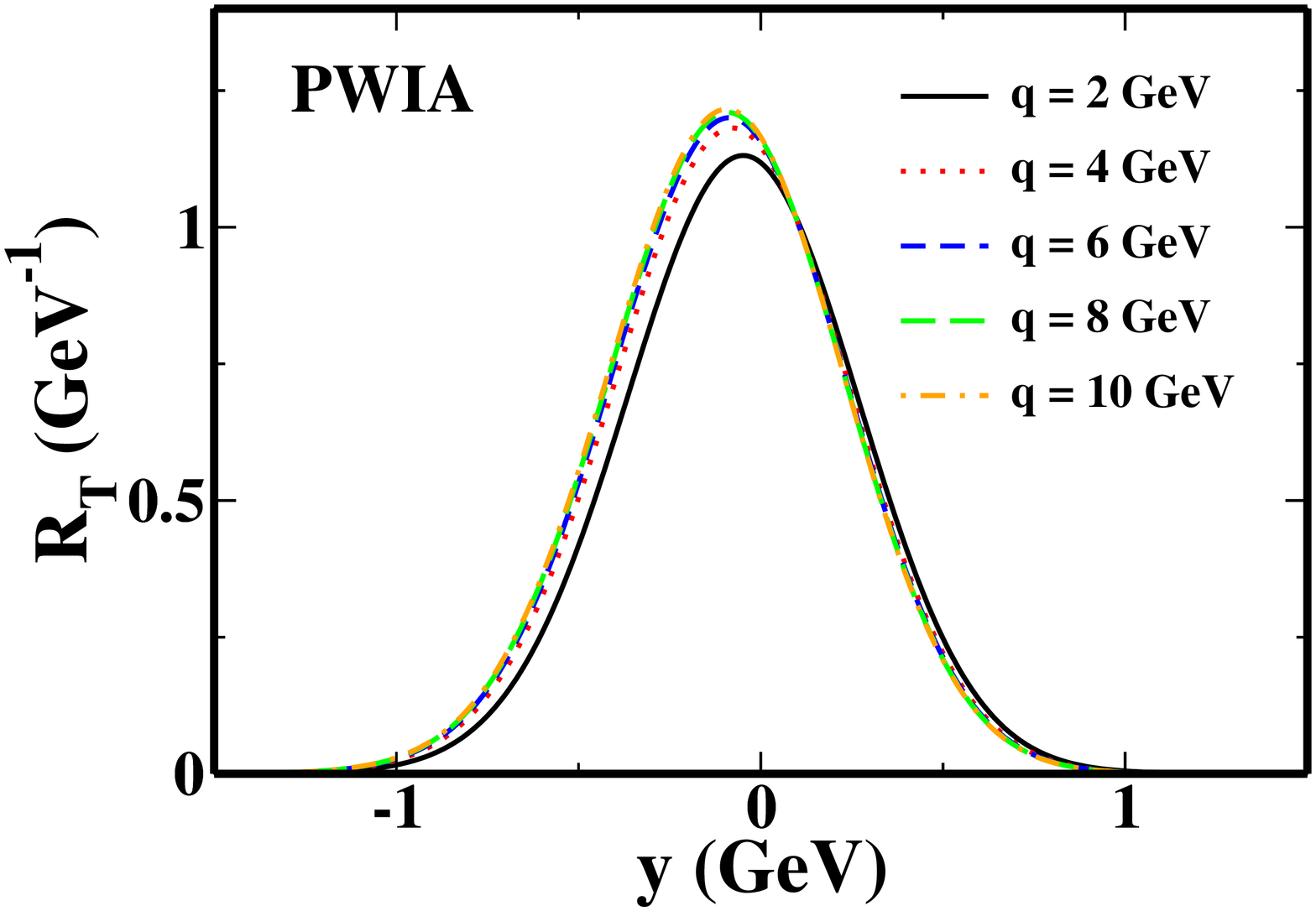}
\caption{The transverse response function $R_T$ is plotted versus
$y$ for several low (bottom panel) and high (top panel) values of
the three-momentum transfer $q$. The results shown have been
calculated in PWIA, using the linear potential.}
\label{figpwlinrt}
\end{figure}

We start by investigating the behavior of the PWIA responses for
the linear potential: $V_s = b r, V_v = 0$. In
Fig.~\ref{figpwlinrl}, bottom panel, we show the longitudinal
response $R_L$ for various, lower three-momentum transfers $q$.
Here, one sees clearly that both the peak height and the width of
the peak are reduced significantly by increasing $q$ for $q <
10~GeV$. The peak position also shifts very slightly to lower $y$
values.  In Fig.~\ref{figpwlinrl}, top panel, we show the
longitudinal response $R_L$ for various, high three-momentum
transfers $q$. There is a small but visible change in going from
$q = 10~GeV$ to $q = 20~GeV$. Scaling is reached for $q = 40~GeV$,
and the curves corresponding to even higher momentum transfers
coincide.

In Fig.~\ref{figpwlinrt}, we show the corresponding results for
the transverse response function $R_T$. For the low momentum
transfers, bottom panel, one sees that increasing $q$ leads to an
increase in peak height and width, along with a small shift of the
peak position towards lower $y$ values. The top panel with the
high momentum transfers shows that scaling sets in faster for
$R_T$ than for $R_L$. Changes are smaller from one $q$ value to
the next, and the curves coincide once $q = 40~GeV$.

\begin{figure}[htp]
\includegraphics[width=20pc,angle=270]{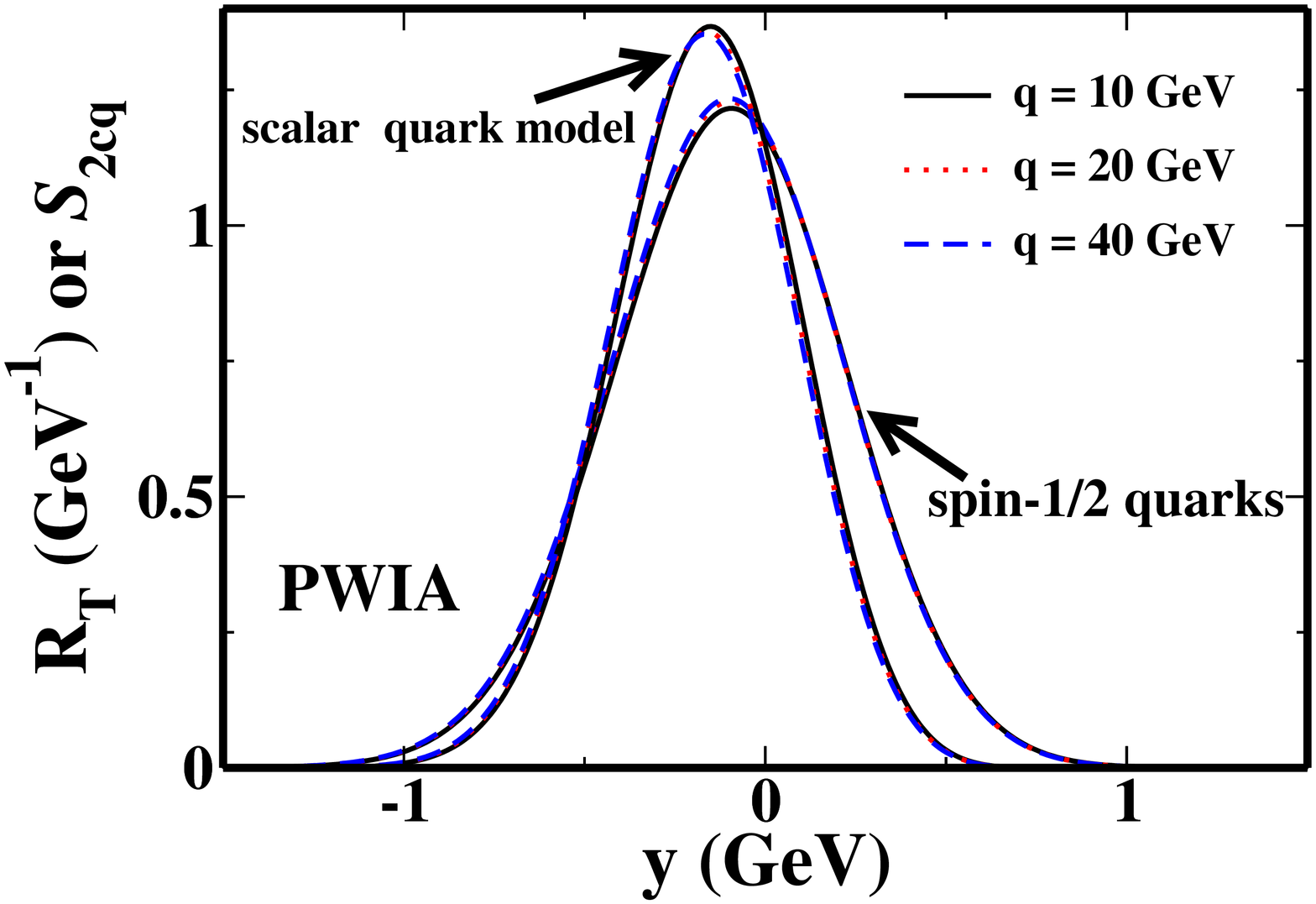}
\caption{A comparison of the onset of scaling for the ``scalar
quark" model and the model with quark spin. The transverse
response function $R_T$ and the scaling function ${\cal{S}}_{2cq}$
of the ``scalar quark" model are plotted versus $y$. The results
shown have been calculated in PWIA, and have already been shown in
Figs.~\ref{figpwlinrt},\ref{figccy}.} \label{figpwspeedofscal}
\end{figure}
It is interesting to consider the onset of scaling when making our
modeling more realistic: in the transition from the ``all scalar"
to the ``scalar quark" model, we observed a considerable slowdown
in the onset of scaling, due to the additional structure, and to
the more complicated form of the terms. Now, we have added the
proper spin to the quarks, but interestingly, we see no
significant change in the scaling behavior. This is illustrated in
Fig.~\ref{figpwspeedofscal}, where we show high $q$ results for
the scalar quark model and our current model. While the two models
obviously lead to different scaling curves, the {\sl onset} of
scaling occurs at roughly the same momentum transfers in both
cases. From this, we have to conclude that the spin of the quark
does not play a significant role in the onset of scaling.

\subsection{y-scaling in FSI}

Now we will show the scaling behavior with FSI included. For
reasons of numerical feasibility, we restrict ourselves to
momentum transfers of $q \leq 10~GeV$.

\begin{figure}
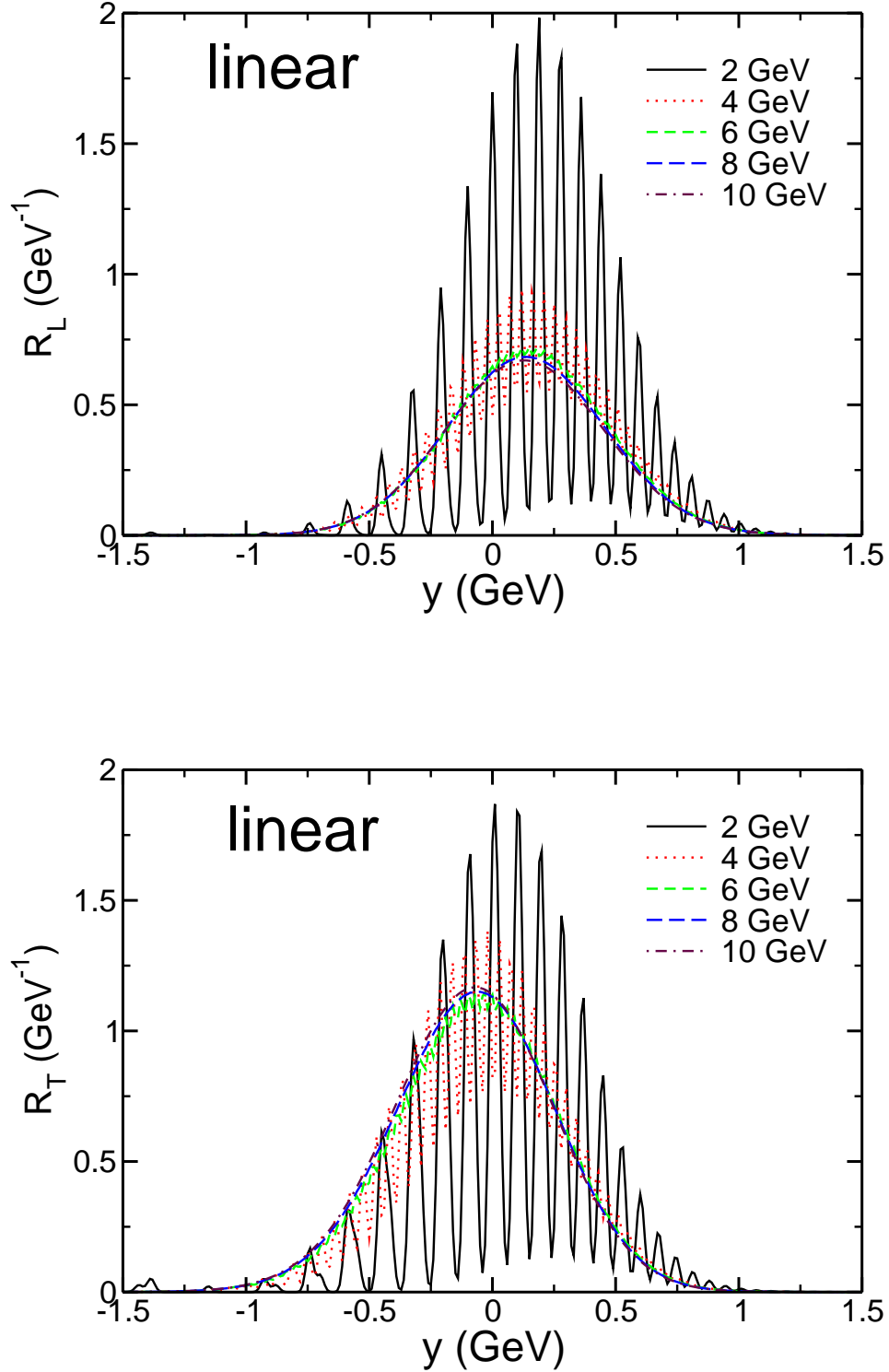

\includegraphics[height=3.5in]{ys_linear_L.eps}

\vspace{2cm}
\includegraphics[height=3.5in]{ys_linear_T.eps}
\caption{The longitudinal response function $R_L$ (top panel) and
the transverse response function $R_T$  are plotted versus $y$ for
several values of the three-momentum transfer $q$. The results
shown have been calculated including FSI, using the linear
potential.}\label{linearResp}
\end{figure}

Figure \ref{linearResp} shows the longitudinal (top panel) and
transverse responses (bottom panel) as a function of $y$ for
momentum transfers from $2$ to $10$ GeV with a width of
$\epsilon=0.02$ GeV for smoothing (recall Eq.~(\ref{smear})).
As the momentum transfer increases, the average response moves to
lower $y$ and since the density of states is also increasing the
curves become increasingly smooth. The longitudinal response
approaches the asymptotic result from above, while the transverse
response approaches it from below. The low $q$ curves with the
visible resonance bumps oscillate around the smoother curves
obtained for higher momentum transfer.

Before comparing the scaling curves obtained in PWIA and with FSI,
we need to investigate the influence of our smoothing procedure
for the delta-function in energy. This is an artifact of our
model, as we assume that the resonances do not decay. In
\cite{jvod2}, we used a Breit-Wigner type smoothing procedure, and
the width $\Gamma$ had some influence on the numerically obtained
scaling curves. However, for the scalar quarks discussed in
\cite{jvod2}, we could find an analytic expression for the scaling
curve with FSI. This is not the case here, so this matter deserves
careful investigation. The PWIA calculation does not suffer from
this problem, as  the knocked-out, ``free" quark can have any
energy, in contrast to the fixed-energy resonance final states
included in FSI. Therefore, comparing the FSI and PWIA scaling
curves is not entirely straightforward.

\begin{figure}[htp]
\includegraphics[width=20pc,angle=270]{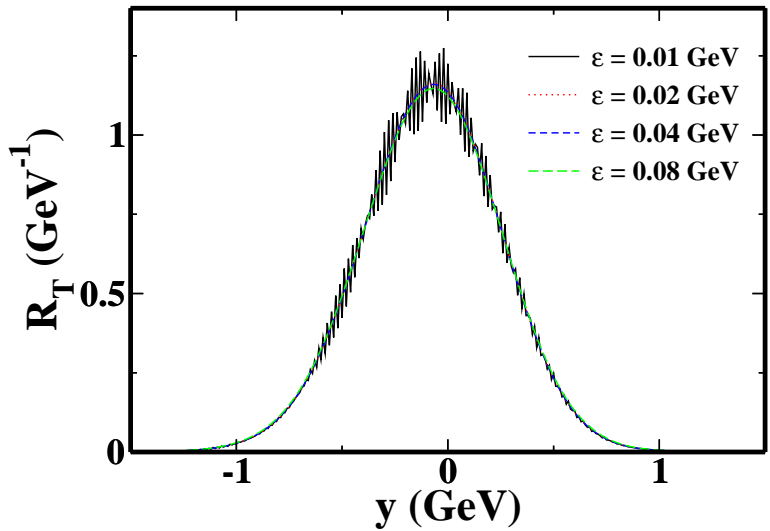}
\caption{The transverse response calculated for the linear
potential in FSI at $q = 10~GeV$ for various values of the
Gaussian smoothing parameter $\varepsilon$.} \label{figsmoothing}
\end{figure}

In Fig.~\ref{figsmoothing}, we show the transverse response $R_T$
at $q = 10~GeV$, including FSI, calculated for various values of
the Gaussian smoothing parameter $\varepsilon$. We show curves for
$\varepsilon = 0.02~GeV$, the value used for all other plots in
this paper, and for $\varepsilon = 0.01, 0.04, 0.08~GeV$. While a
smaller value of $\varepsilon$ leads to less smoothing and visible
resonance bumps even at higher $q$, the overall shape, peak
position and width are not changed. For larger values of
$\varepsilon$, there is no visible difference to the original
curve with $\varepsilon = 0.02~GeV$ . Therefore, the dependence on
the smoothing parameter $\varepsilon$ is so weak that it will not
influence our comparison of the PWIA and FSI scaling curves.

\subsection{Model dependence}

\begin{figure}
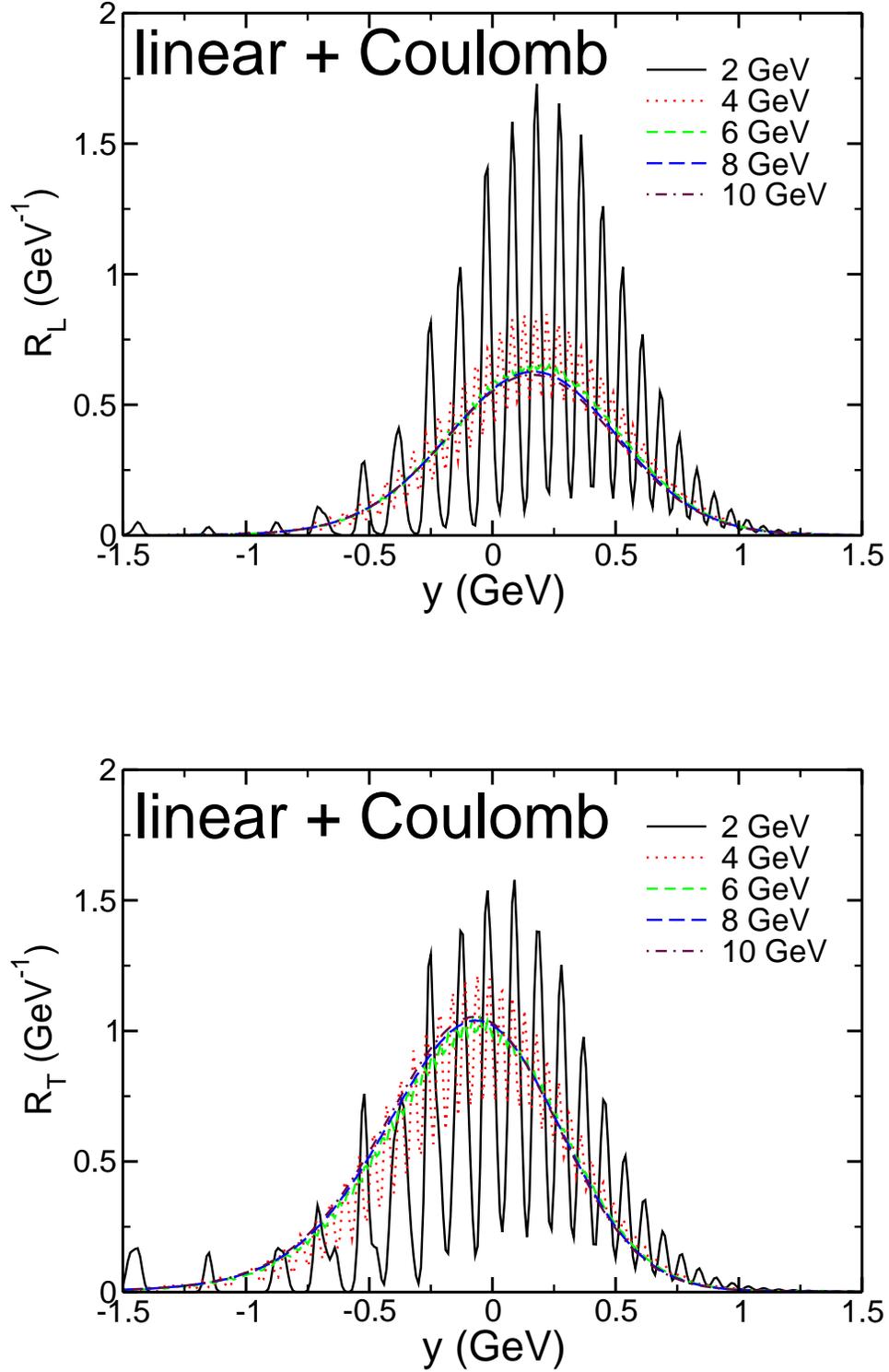

\centerline{\includegraphics[height=3.5in]{ys_coulomb_L.eps}}

\vspace{2cm}
\centerline{\includegraphics[height=3.5in]{ys_coulomb_T.eps}}
\caption{The longitudinal response function $R_L$ (top panel) and
the transverse response function $R_T$  are plotted versus $y$ for
several values of the three-momentum transfer $q$. The results
shown have been calculated including FSI, using the
linear-plus-Coulomb potential.}\label{coulombResp}
\end{figure}

\begin{figure}
\centerline{\includegraphics[height=3.5in]{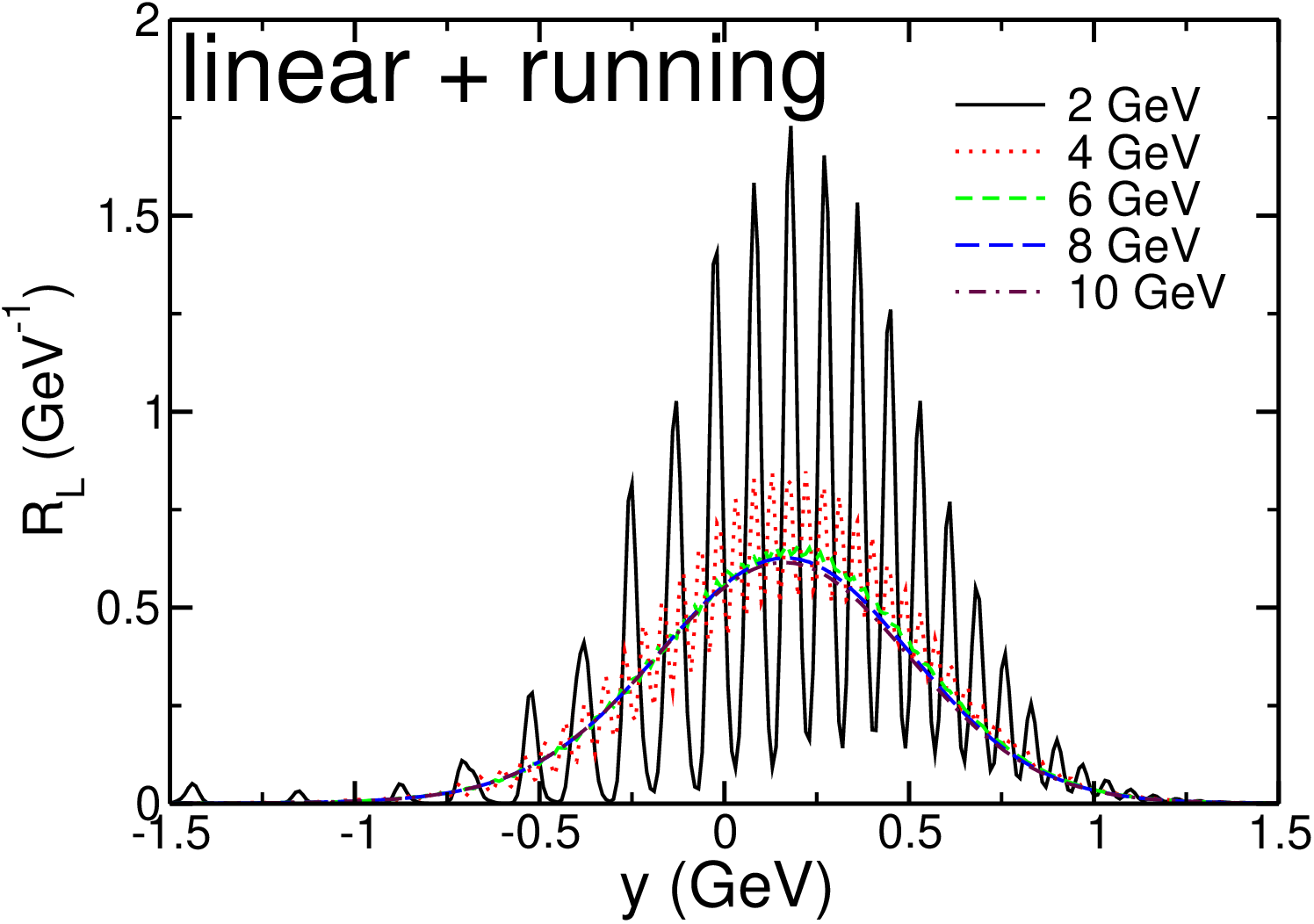}}

\vspace{2cm}
\centerline{\includegraphics[height=3.5in]{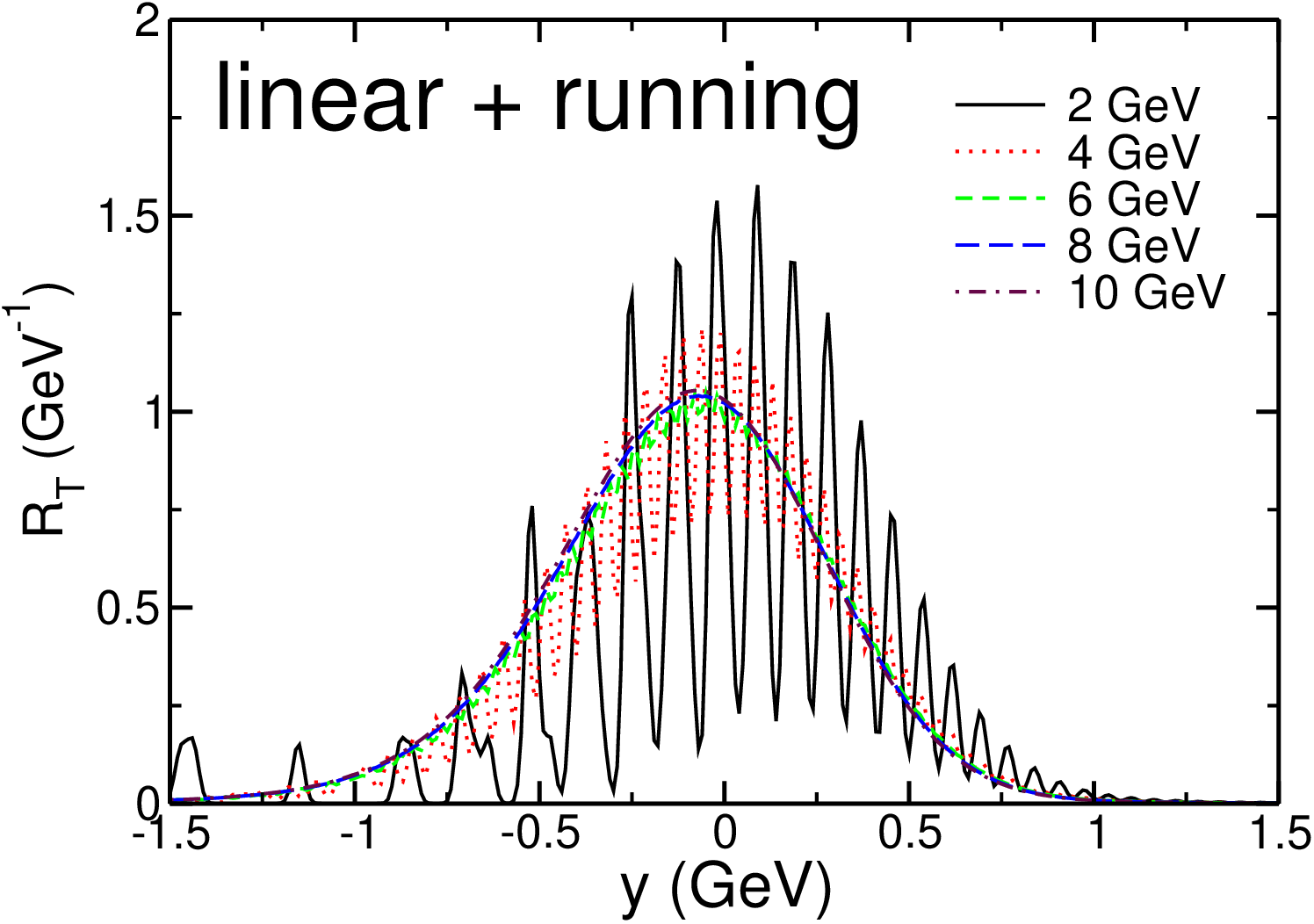}}
\caption{The longitudinal response function $R_L$ (top panel) and
the transverse response function $R_T$  are plotted versus $y$ for
several values of the three-momentum transfer $q$. The results
shown have been calculated including FSI, using the
linear-plus-running potential.}\label{runningResp}
\end{figure}

Calculations for the linear-plus-coulomb  potential are shown in
Fig. \ref{coulombResp}, and for the linear-plus-running potential
in Fig. \ref{runningResp}. These figures show features consistent
with those of the linear potential alone. One can see that the
linear potential leads to the highest peak value for both $R_T$
and $R_L$, and has the narrowest width, while the
linear-plus-running potential has the lowest peak height and the
largest width. In all cases, we have not quite reached the scaling
limit yet, but as the changes from $q = 6~GeV$ to $q = 10~GeV$ are
small, we are not far away from it, either. The low $q$ results
clearly oscillate around the high $q$ results, as expected from
duality.

\begin{figure}
\centerline{\includegraphics[height=4in]{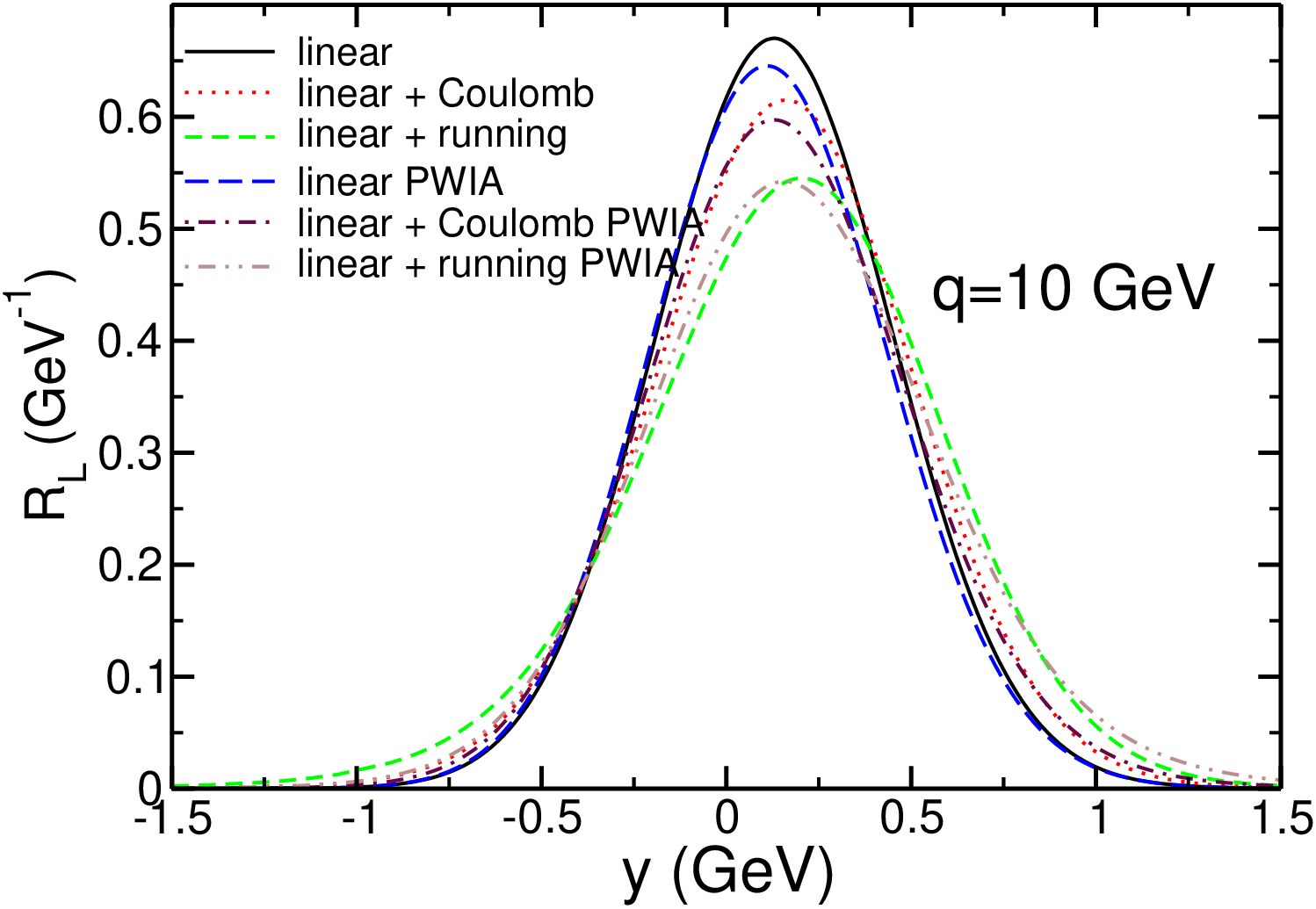}}

\vspace{2cm}
\centerline{\includegraphics[height=4in]{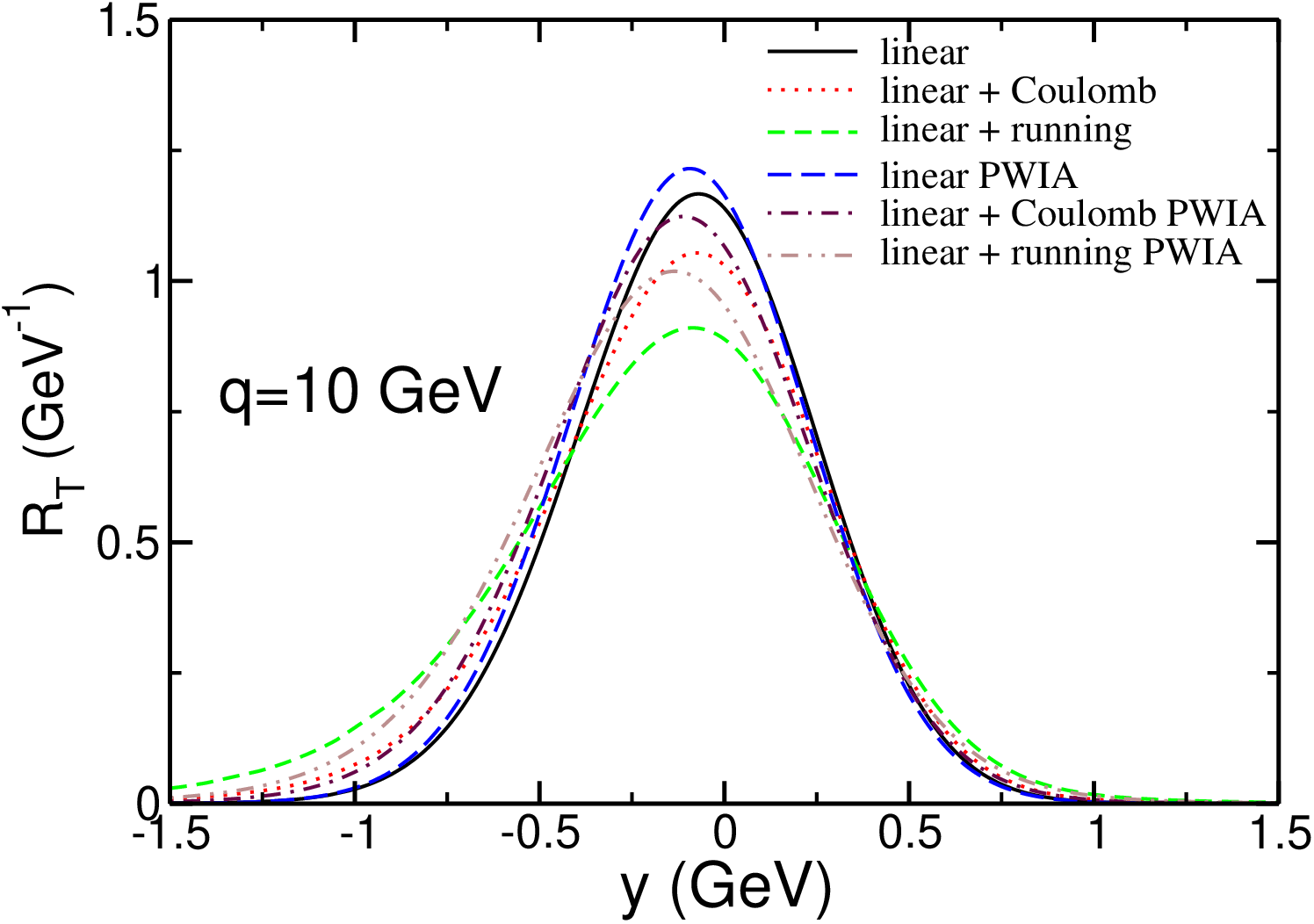}}
\caption{}\label{summaryResp}
\end{figure}

Figure \ref{summaryResp} shows a comparison of the FSI and PWIA
calculations at $q=10$ GeV for the longitudinal and transverse
responses. At this momentum transfer the corresponding FSI and
PWIA calculations differ by small changes in magnitude and peak
position. Since the calculations are not yet converged to the
scaling limit, it is not clear whether this represents a failure
to scale to the same limit or is simply a manifestation of
differing rates of convergence.

On a fundamental level, this is important as FSI is usually
assumed to be negligible in the analysis of deep inelastic
scattering data. Recently, however, some authors have pointed out
that FSI - through gluon exchange between fast, outgoing partons
and target spectators - can make contributions to the leading
twist structure functions at small $x_{Bj}$ \cite{hoyer}. Still,
the changes we see here are small, and one does not make much of a
mistake in neglecting the effect of all the FSIs combined.

\section{Sum Rules}

Some of the features of this model can be explored further and
more precisely by means of energy-weighted sum rules. This will
allow us to get a better feeling for the bulk features of our
results, like peak position and width. Moments that become
constant at high momentum transfer are another signature of
quark-hadron duality \cite{prdmom}, and with the sum rules derived
in this section we can test if our model behaves in this way.

By using an integral representation of the energy conserving delta
function and replacing eigenenergies with the hamiltonian
operator, the response functions can be written as
\begin{eqnarray}
R_L(q,\nu)&=&\int_{-\infty}^\infty\frac{dt}{2\pi}e^{i\nu t}
<\Psi_0|e^{iH(\hat{\bm{p}},\bm{x})
t}e^{-i\bm{q}\cdot\bm{x}}e^{-iH(\hat{\bm{p}},\bm{x}) t}
e^{i\bm{q}\cdot\bm{x}}|\Psi_0>\nonumber\\
&=&\int_{-\infty}^\infty\frac{dt}{2\pi}e^{i\nu t} <\Psi_0|e^{i\hat
H(\hat{\bm{p}},\bm{x}) t}e^{-iH(\hat{\bm{p}}+\bm{q},\bm{x}) t}
|\Psi_0>
\end{eqnarray}
and
\begin{eqnarray}
R_T(q,\nu)&=&\sum^2_{i=1}\int_{-\infty}^\infty\frac{dt}{2\pi}e^{i\nu
t} <\Psi_0|e^{i H(\hat{\bm{p}},\bm{x})
t}\alpha_ie^{-i\bm{q}\cdot\bm{x}}e^{-iH(\hat{\bm{p}},\bm{x}) t}
e^{i\bm{q}\cdot\bm{x}}\alpha_i|\Psi_0>\nonumber\\
&=&\sum^2_{i=1}\int_{-\infty}^\infty\frac{dt}{2\pi}e^{i\nu t}
<\Psi_0|e^{i\hat H(\hat{\bm{p}},\bm{x})
t}\alpha_ie^{-iH(\hat{\bm{p}}+\bm{q},\bm{x}) t}\alpha_i
|\Psi_0>\,,
\end{eqnarray}
where the momentum shift operator has been recognized and used in
the last step for both response functions.

The energy weighted sum rules are defined as integrals of the
response functions over $-\infty<\nu<\infty$ weighted with powers
of $\nu$. This integral can be simplified by writing
\begin{equation}
\nu^ne^{i\nu t}=\left(-i\frac{\partial}{\partial t}\right)^n
e^{i\nu t}
\end{equation}
and then integrating by parts to give
\begin{eqnarray}
{\cal S}_{Ln}(q)&=&\int_{-\infty}^\infty d\nu\,
\nu^n R_L(q,\nu)\nonumber\\
&=&\left(i\frac{\partial}{\partial t}\right)^n
<\Psi_0|e^{iH(\hat{\bm{p}},\bm{x})
t}e^{-iH(\hat{\bm{p}}+\bm{q},\bm{x}) t} |\Psi_0>_{t=0}
\end{eqnarray}
and
\begin{eqnarray}
{\cal S}_{Tn}(q)&=&\int_{-\infty}^\infty d\nu\,
\nu^n R_T(q,\nu)\nonumber\\
&=&\sum^2_{i=1}\left(i\frac{\partial}{\partial t}\right)^n
<\Psi_0|e^{iH(\hat{\bm{p}},\bm{x})
t}\alpha_ie^{-iH(\hat{\bm{p}}+\bm{q},\bm{x}) t}\alpha_i
|\Psi_0>_{t=0}\,.
\end{eqnarray}
The momentum-shifted hamiltonian for our model is
\begin{equation}
H(\hat{\bm{p}}+\bm{q},\bm{x})=\bm{\alpha}\cdot\bm{q}+H(\hat{\bm{p}},\bm{x})\,.
\end{equation}
As a result, the sum rules can be reduced to nested commutators
involving $H$, $\bm{\alpha}\cdot\bm{q}$ and $\alpha_i$.

The first three sum rules for the longitudinal response are
\begin{eqnarray}
{\cal S}_{L0}(q)&=&<\psi_0|\psi_0>=1\nonumber\\
{\cal S}_{L1}(q)&=&<\psi_0|\bm{\alpha}\cdot\bm{q}|\psi_0>=0\nonumber\\
{\cal S}_{L2}(q)&=&q^2
\end{eqnarray}
The first of these is the Coulomb sum rule and simply reflects the
conservation of charge. We have assumed that the charge of the
bound quark is 1 for simplicity. The second of these indicates
that the response is roughly antisymmetric about $\nu=0$. The
third indicates that the width of the distribution is increasing
as $q$.  This is consistent with the case where the longitudinal
response consists of peaks at positive and negative energy
transfers of roughly equal magnitude and opposite signs that are
moving away from the origin in opposite directions by a distance
proportional to $q$. The existence of the negative energy peak is
a consequence of the presence of the negative energy solutions to
the Dirac equation which are not physical in a simple one-body
model.

The positive energy contribution can be isolated by defining the
positive energy moments
\begin{equation}
S_{L(T)n}(q)=\int_{0}^\infty d\nu\, \nu^nR_{L(T)}(\bm{q},\nu)\,.
\end{equation}
Since the $n=0$ moment is no longer normalized to one, we can
define the average energy transfer as
\begin{equation}
\left< \nu\right>_{L(T)}=\frac{S_{L(T)1}}{S_{L(T)0}}
\end{equation}
and the rms variation in energy transfer as
\begin{equation}
\Delta\nu_{L(T)}=\sqrt{\frac{S_{L(T)2}}{S_{L(T)0}}-\left<
\nu\right>_{L(T)}^2}\,.
\end{equation}

\begin{figure}
\centerline{\includegraphics[height=4in]{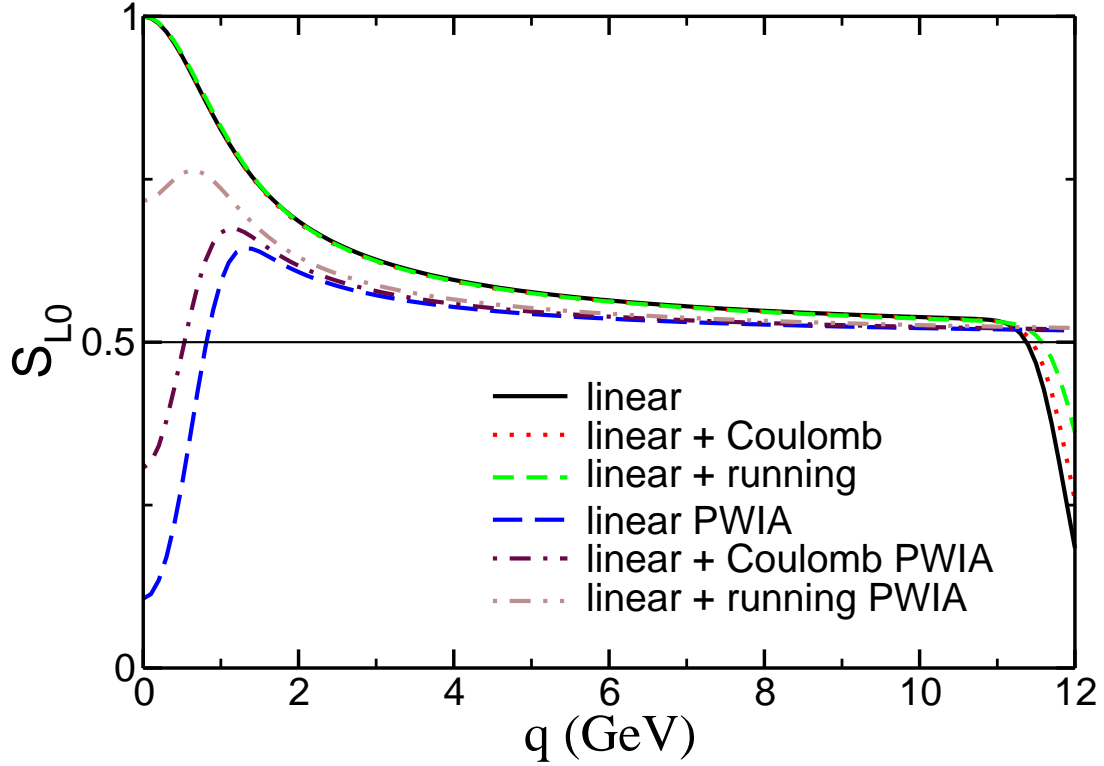}} \caption{The
positive energy $n=0$ moment of the  longitudinal response
function, calculated with FSI and in PWIA  for the linear,
linear-plus-coulomb, and linear-plus-running
potentials.}\label{sl0}
\end{figure}

\begin{figure}
\centerline{\includegraphics[height=4in]{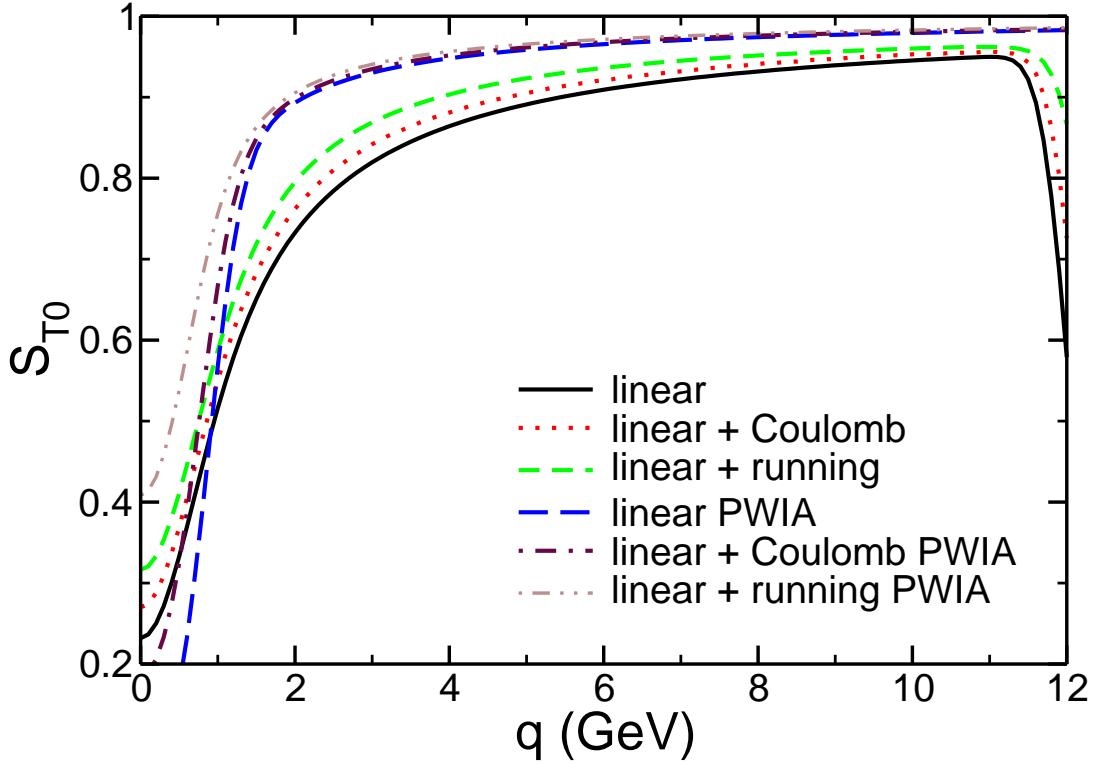}} \caption{The
positive energy $n=0$ moment of the transverse response function,
calculated with FSI and in PWIA  for the linear,
linear-plus-coulomb, and linear-plus-running
potentials.}\label{st0}
\end{figure}

Figures \ref{sl0} and \ref{st0} show the positive energy $n=0$
moments of the  longitudinal and transverse response functions.
For $S_{L0}$, the results for the linear, linear-plus-coulomb, and
linear-plus-running potentials are virtually identical except at
$q>10\ {\rm GeV}$ where the moment starts to fall off due to
finite range of excitation energies that have been calculated. The
corresponding calculations in PWIA vary at small $q$ due to
variations in the available phase space, but must approach
$\frac{1}{2}$ for $q\rightarrow\infty$. It is clear from this
figure that neither the FSI calculations nor the PWIA calculations
have reached their saturation values at the highest values of $q$
that have been calculated here. Although it is plausible that the
full results may be approaching the PWIA values for large $q$, it
is not possible to determine that this is the case based on these
calculations.

Similar results can be seen for the transverse moment shown in
Fig. \ref{st0} although there is a greater variation among the
full calculations at lower values of $q$.  In this case the
asymptotic value of the moment for the PWIA calculations is 1.

\begin{figure}
\centerline{\includegraphics[height=4in]{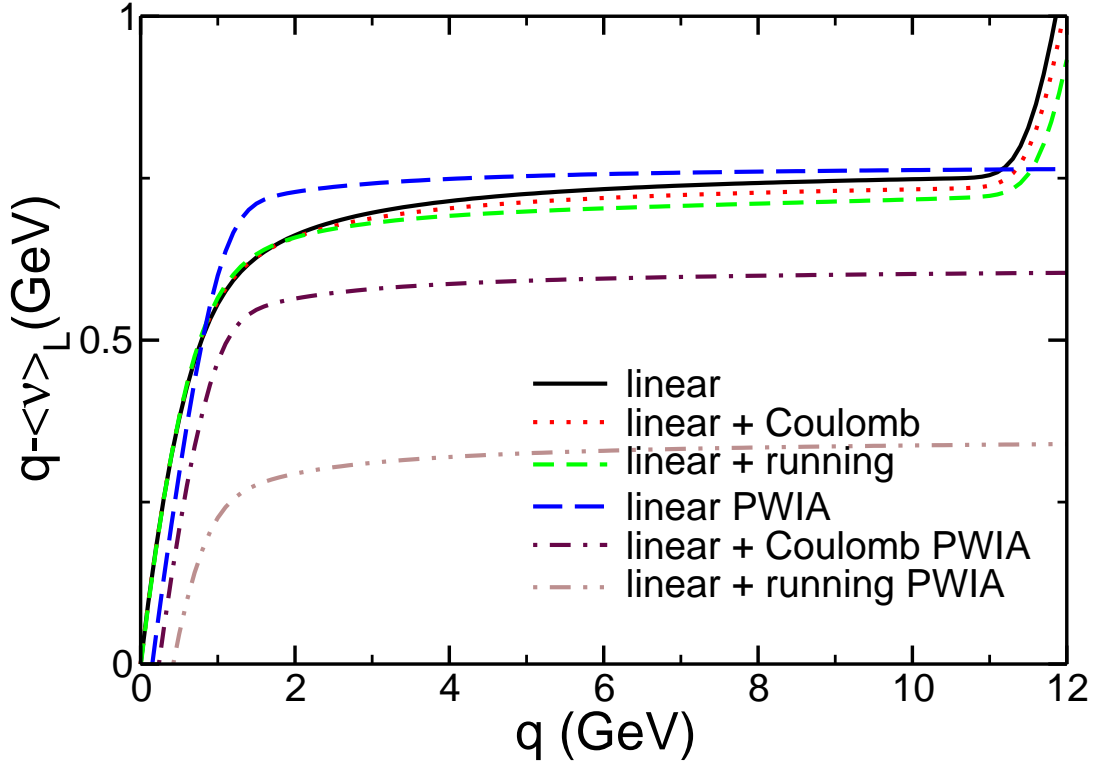}} \caption{The
difference $q-\left<\nu\right>_L$ between the average energy
transfer $\left<\nu\right>_L$ and the momentum transfer for the
longitudinal response, calculated with FSI and in PWIA  for the
linear, linear-plus-coulomb, and linear-plus-running potentials.}
\label{qMnuL}
\end{figure}

The asymptotic values of $\left<\nu\right>_{L(T)}$ for the PWIA
are given by
\begin{equation}
\lim_{q\rightarrow\infty}\left<\nu\right>_L=q-E_0+\frac{1}{12\pi^2}
\int_0^\infty dp\,p^3 n_v^s(p)
\end{equation}
and
\begin{equation}
\lim_{q\rightarrow\infty}\left<\nu\right>_T=q-E_0-\frac{1}{12\pi^2}
\int_0^\infty dp\,p^3 n_v^s(p)\,.
\end{equation}
Note that the longitudinal and transverse responses are offset
from one another by terms that depend upon $n_v^s(p)$. Since the
average value of the energy transfer increases linearly with $q$,
it is convenient to plot this moment as $q-\left<\nu\right>_L$.
This quantity is shown in Fig. \ref{qMnuL} for the three FSI and
three PWIA calculations used in the previous figures. Here the
three FSI calculations have very similar average positions while
the average positions of the PWIA calculations show large
differences. This is the result of the sensitivity of the position
of the peak of the PWIA response to the difference in energy of
the bound state and the lowest plane-wave state. In the FSI
calculation, both the ground and excited states see the same
potential which seems to limit the size of the shift in peak
position for the three different potentials.
For purposes of comparison, it is possible to eliminate this
disparity in the PWIA calculations by introducing a constant
vector potential into the Dirac equation. This has the effect of
simply shifting the spectrum. As a result we can use this shift to
place all of the ground state energies at the same value. This
shift has no effect on the response functions for the FSI
calculations, but will correct for the differences in phase space
in the PWIA. In fact, we already have applied this shift above in
the section on y-scaling.

\begin{figure}
\centerline{\includegraphics[height=4in]{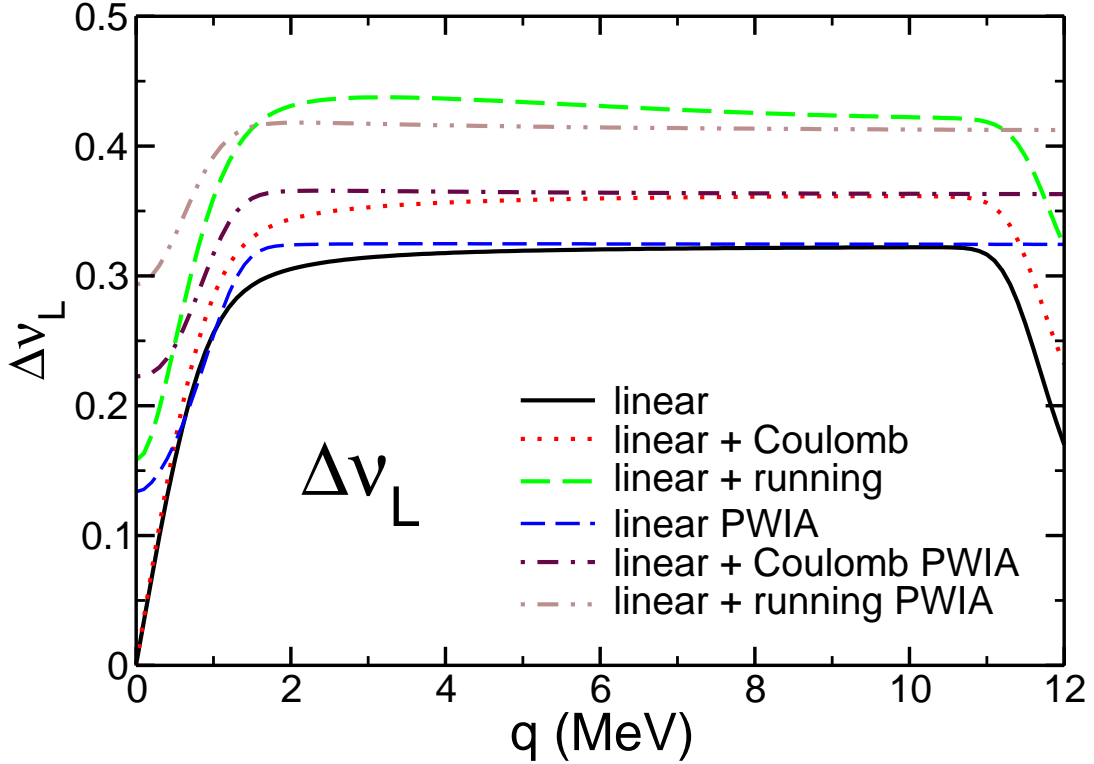}}
\caption{The rms width $\Delta \nu_L$ for the longitudinal
response, calculated with FSI and in PWIA  for the linear,
linear-plus-coulomb, and linear-plus-running potentials.}
\label{deltanul}
\end{figure}

Figure \ref{deltanul} shows the rms widths for the various
calculations of the longitudinal response. The asymptotic value of
this width in the PWIA is given by
\begin{equation}
\lim_{q\rightarrow\infty}\Delta\nu_L=\left[\frac{1}{12\pi^2}\int_0^\infty
dp\, p^4 n_v^0(p)- \frac{1}{144\pi^4}\left(\int_0^\infty dp\, p^3
n_v^s(p)\right)^2\right]^\frac{1}{2}\,.
\end{equation}
In all three cases the width of the FSI calculation is close to
that of the corresponding PWIA calculation.  This suggests that
the width of the response is determined largely by the width of
ground state momentum distributions.

While we have not yet reached convergence to the scaling limit at
the momentum transfers for which we have calculated, one can
clearly see that the moments do flatten out with higher momentum
transfers, qualitatively agreeing with the observations in duality
experiments.

\section{The ``y-scaling Callan-Gross relation"}

From $x$-scaling in the deep inelastic region, the Callan-Gross
relation \cite{callangross} $2 x F_1 (x) = F_2 (x)$ is known to
hold. The physical significance of the Callan-Gross relation is
that one scatters off spin $1/2$ objects, which have a dominant
contribution from the magnetization current, i.e. in the
transverse part of the structure functions. In the real world, the
Callan-Gross relation needs to be corrected for radiative effects,
and is then observed to hold in the experimental data from the
deep inelastic region. In our model, there are no radiative
corrections, as we do not allow the production of new particles.
Our model is, so far, entirely quantum-mechanical in this sense.
Therefore, we expect the Callan-Gross relation to hold as is in
$x$ or $u$ scaling. As we focus on $y$ scaling, we now derive the
analog of the Callan-Gross relation, using the fact that both the
longitudinal and the transverse response scale for large $q$.

The structure functions $W_1, W_2$ are related to the longitudinal
and transverse responses $R_L$ and $R_T$ by
\begin{equation}
W_1 = \frac{1}{2} R_T
\end{equation}
and
\begin{equation}
W_2 = \frac{Q^4}{q^4} R_L + \frac{Q^2}{2 q^2} R_T \,
\end{equation}
The Callan-Gross relation can be obtained from these equations by
re-expressing $q$ and $\nu$ in terms of $x$ and $Q^2$, and
increasing $Q^2$ at fixed $x$. In order to find an analogous
relation for $y$ scaling, we re-express $Q^2$ and $\nu$ in terms
of $q$ and $y$, and then increase $q$ at fixed $y$. The kinematic
factors in this limit become
\begin{equation}
 \nu  \frac{Q^2}{2 q^2} \to E_0 - y
\end{equation}
and

\begin{equation}
  \frac{Q^4}{q^4} \to O \left( \frac{1}{q^2} \right ) \to 0 \,.
\end{equation}

The two functions, $F_1 = M W_1$ and $F_2 = \nu W_2$, in this
limit are:

\begin{equation}\label{f1lim}
  F_1 = M W_1 = \frac{1}{2} M  R_T \to \frac{1}{2} M  R_T
\end{equation}

\begin{equation}\label{f2lim}
  F_2 = \nu W_2 = \nu \frac{Q^4}{q^4} R_L + \nu \frac{Q^2}{2 q^2}
  R_T \to (E_0 - y) R_T
\end{equation}

Direct comparison of Eqs. (\ref{f1lim}) and (\ref{f2lim}) yields
$F_2 = F_1 \frac{2}{M} (E_0 - y)$. Substituting the definitions of
$F_1$ and $F_2$, one obtains the analog of the Callan-Gross
relation for $y$-scaling:

\begin{equation}\label{cgy}
  2 W_1 (E_0 - y) = \nu W_2 \, .
\end{equation}

\begin{figure}[ht]
\includegraphics[width=20pc,angle=270]{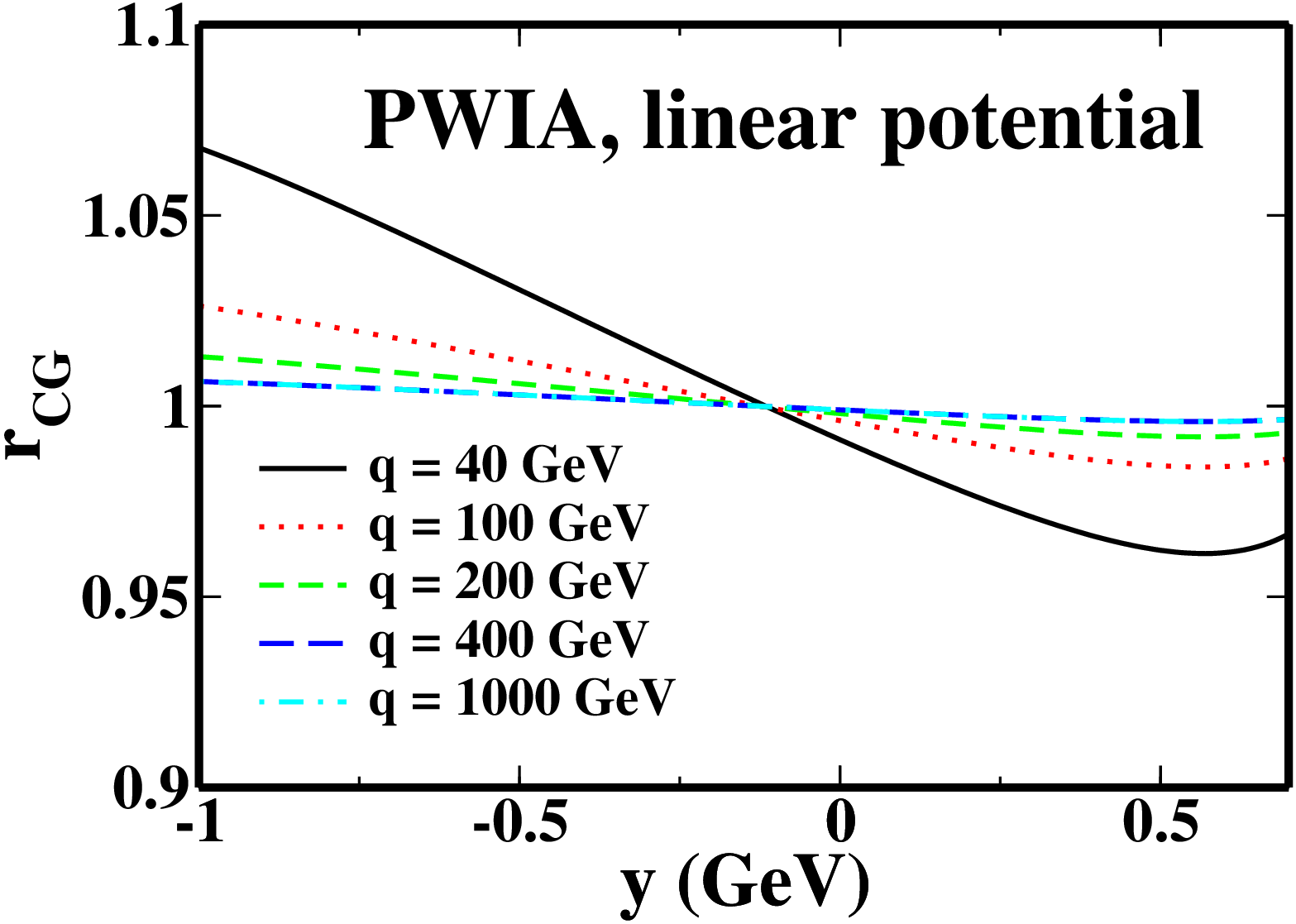}
\includegraphics[width=20pc,angle=270]{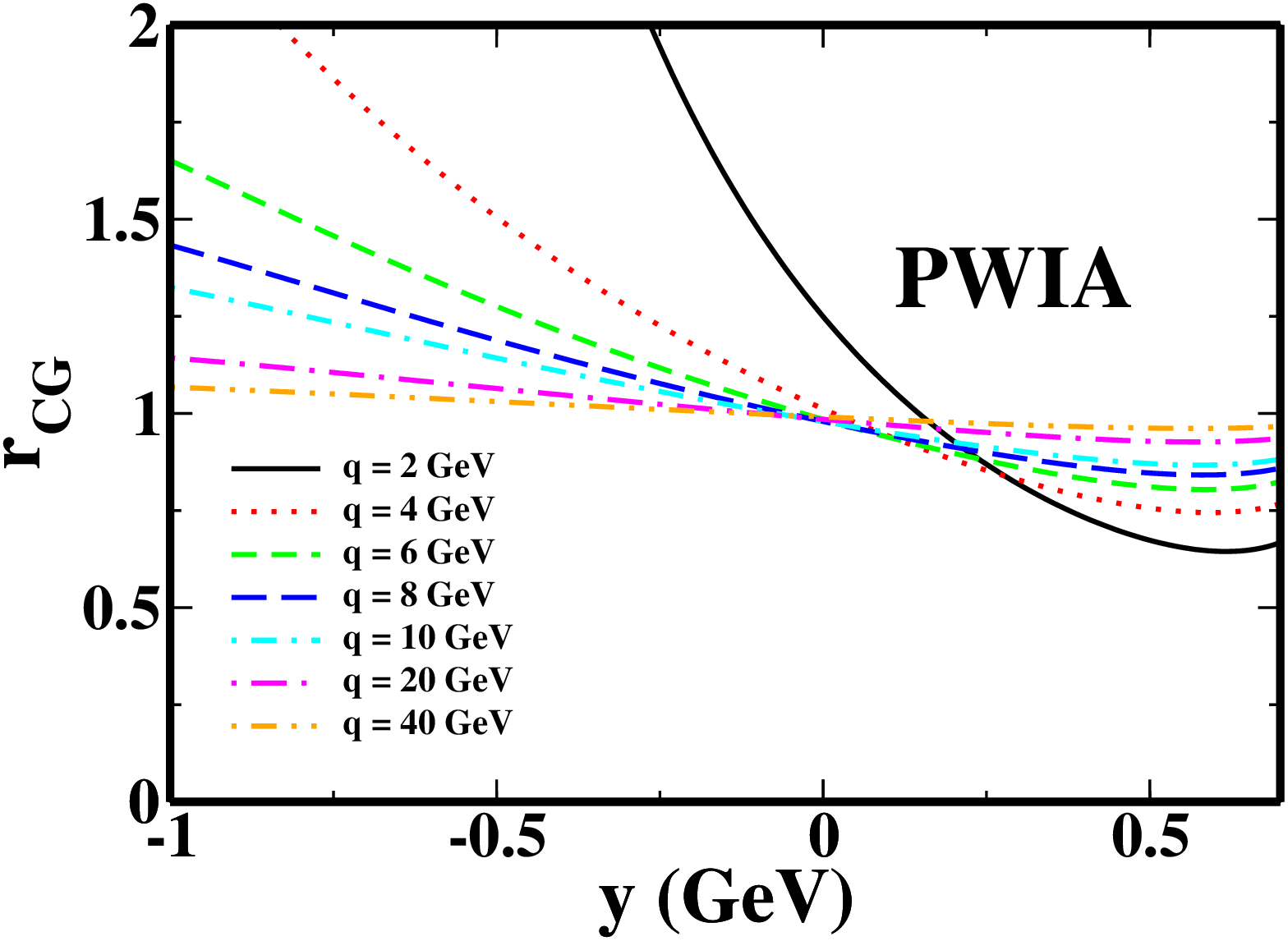}
\caption{The Callan-Gross ratio $r_{CG}$ is plotted versus $y$ for
several values of the three-momentum transfer $q$. The results
shown have been calculated in PWIA with a linear potential. Note
the different scales employed in the two panels.}
\label{figcgpwlinhiq}
\end{figure}

We now proceed to check if our numerical results fulfill the
Callan-Gross relation, i.e. if we have reached the scaling region
already. First, we show results for PWIA, where we can access
arbitrarily high $q$ without any numerical problems. In order to
display the approach to scaling, we show the ratio of the
left-hand-side to the right-hand-side of Eq. (\ref{cgy}), $r_{CG}
= 2 W_1 (E_0 - y)/ \nu W_2$. In Fig.~\ref{figcgpwlinhiq}, top
panel, we show $r_{CG}$ for the linear potential for values of the
three momentum ranging from $q = 10~GeV$ to $q = 1000~GeV$. As
expected, scaling has set in at these high momentum transfers: the
deviations of $r_{CG}$ from $1$ are very small, at the level of $
5 \%$ or less for $q \geq 60~ GeV$. It is interesting to note that
for higher $y$ values, $\nu W_2$ dominates, while for lower $y$
values, $2 W_1 (E_0 - y)$ is larger. As in the calculation of the
response functions, we have applied the energy shift to adjust the
difference in phase space in PWIA. Note that we have omitted the
regions $y < -1 GeV$ and $y > 0.7 GeV$ from our graphs, as they do
not contain any strength in the responses. The latter is the area
where $(E_0 - y)$ undergoes a sign change, leading to a spike in
the ratio $r_{CG}$ that is unrelated to the interesting physics -
there is hardly any strength in the responses for $y > 0.5~GeV$.

For lower $q$ values, see bottom panel of
Fig.~\ref{figcgpwlinhiq}, scaling has clearly not yet set in, and
the deviations of $r_{CG}$ from unity are large. Comparing these
results for $r_{CG}$ with the onset of scaling observed in the
PWIA response functions $R_L$ and $R_T$, as discussed in Section
\ref{secpwiayscal}, we see that $r_{CG}$ gives additional
information on the scaling behavior. While the response function
plots show that the curves coincide for $q = 40 ~GeV$, at the
latest, the Callan-Gross ratio is more sensitive and still shows
slight changes for increasing three-momentum transfers.

\begin{figure}[ht]
\includegraphics[width=20pc,angle=270]{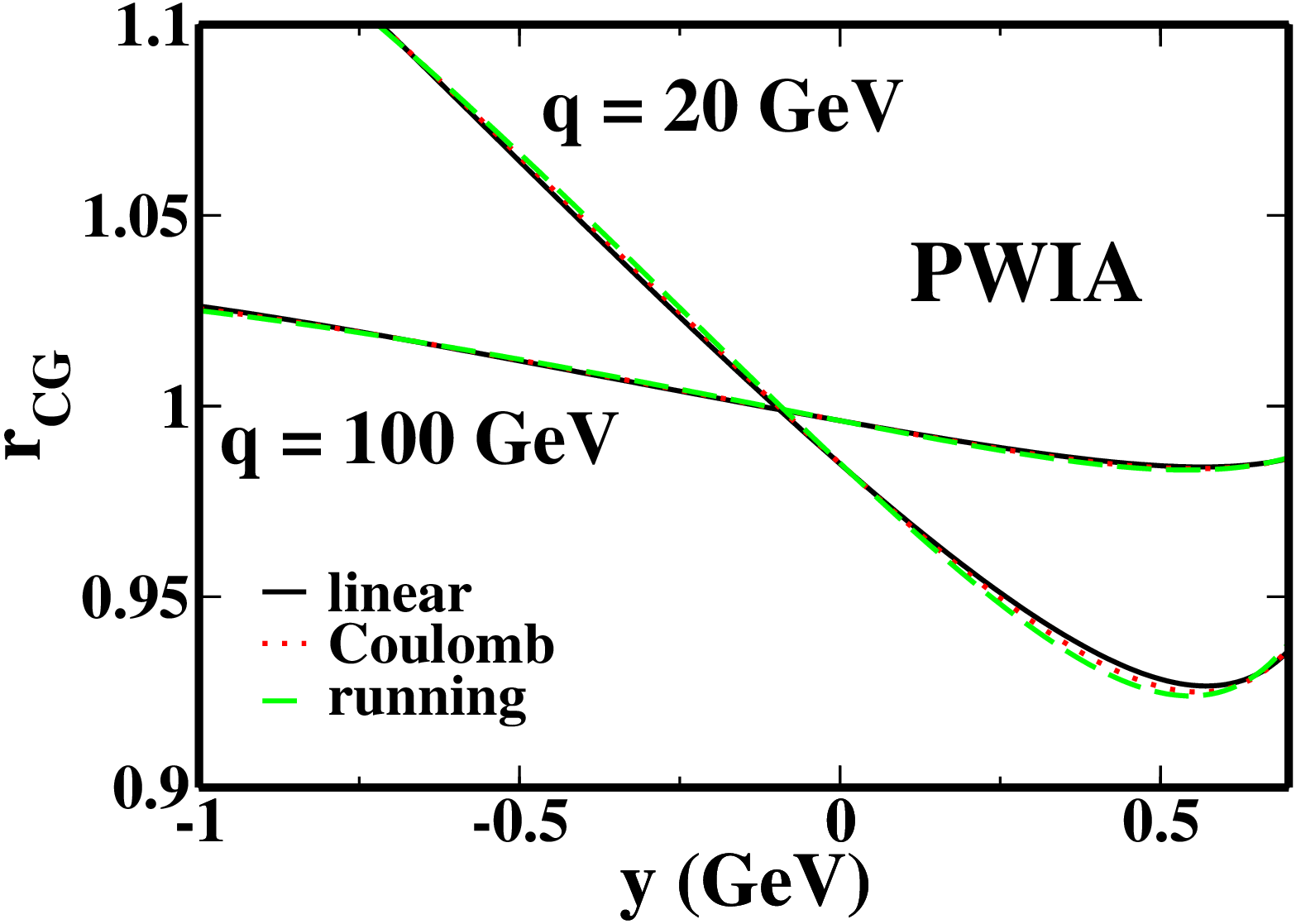}
\includegraphics[width=20pc,angle=270]{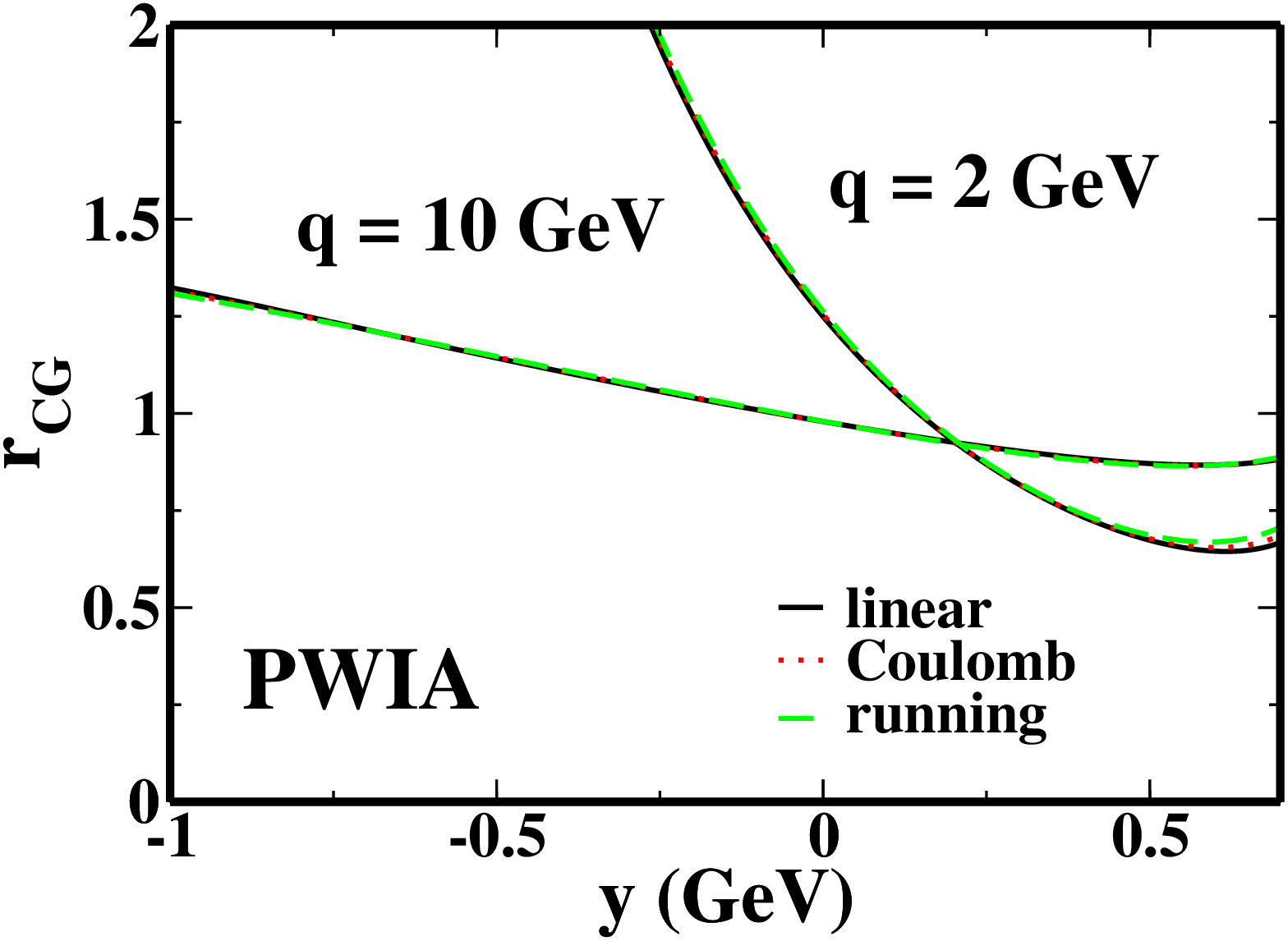}
\caption{The Callan-Gross ratio $r_{CG}$ is plotted versus $y$ for
the linear, Coulomb and running potential, for three-momentum
transfers $q = 2, 10, 20, and 100 GeV$. The results shown have
been calculated in PWIA. Note the different scales employed in the
two panels.} \label{figcgpwpot}
\end{figure}
Comparing the approach of $r_{CG}$ to $1$ for different
potentials, Fig.~\ref{figcgpwpot} shows that linear, Coulomb, and
running potentials have an almost identical scaling behavior, both
at low and high $q$ values.

\begin{figure}[ht]
\includegraphics[width=20pc,angle=270]{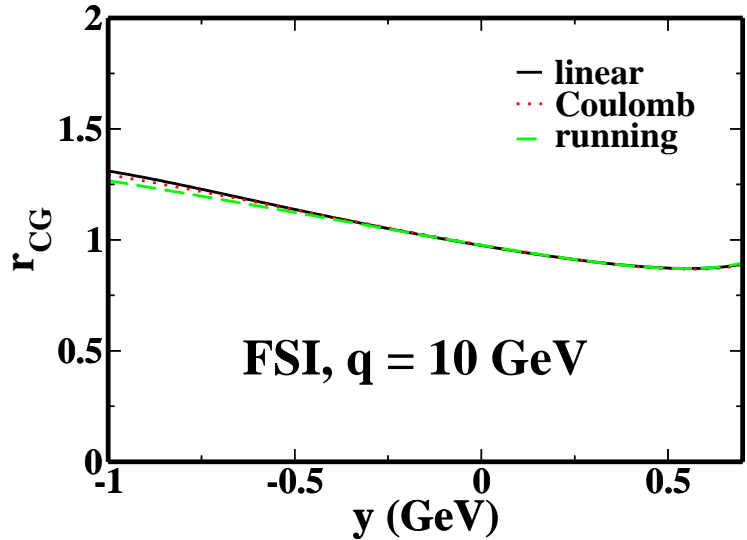}
\caption{The Callan-Gross ratio $r_{CG}$ is plotted versus $y$ for
 three-momentum transfer $q = 10~GeV$. The results
shown have been calculated in FSI with a linear, Coulomb, and
running potential.} \label{figcgfsi}
\end{figure}

In Fig.~\ref{figcgfsi}, we show $r_{CG}$ calculated including FSI
for $q = 10~GeV$, the highest value we have attained for FSI
calculations. The calculations shown are performed for the
different potentials considered in this paper, and one can see
that, just like for PWIA, the results for the Callan-Gross ratio
$r_{CG}$ are very similar.

\begin{figure}[ht]
\includegraphics[width=20pc,angle=270]{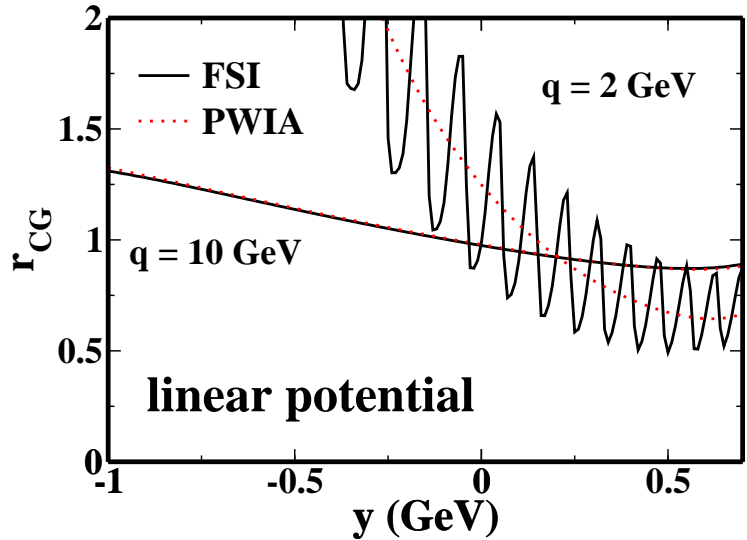}
\caption{The Callan-Gross ratio $r_{CG}$ is plotted versus $y$ for
 three-momentum transfers $q = 2~GeV$ and $q = 10~GeV$. The results
shown have been calculated in PWIA and FSI with a linear
potential.} \label{figcgpwfsi}
\end{figure}

This similarity to the behavior of the PWIA curves leads us to a
direct comparison of the PWIA and FSI results for the linear
potential in Fig.~\ref{figcgpwfsi}. The PWIA and FSI agree very
nicely. For lower $q$ values, the FSI results obviously lead to
oscillatory behavior, but they average to the smooth PWIA result.
The PWIA and FSI results for the Coulomb potential also track each
other very closely. For the running potential, the differences
between FSI and PWIA are a bit more visible, but FSI and PWIA
results are still very close.

While this is no proof, this behavior leads one to the conjecture
that scaling for FSI should set in at the same $q$ value as for
PWIA, and that the {\sl rate of convergence} in both cases is the
same. Note that this result does not imply that the individual
scaling results for PWIA and FSI are the same. Quite on the
contrary, as we found that the FSI results at $q = 10~GeV$ are not
that close to the PWIA scaling curve, and as the PWIA results did
not change dramatically from $q = 10~GeV$ to higher $q$ values,
the conjecture of the same rate of convergence suggests that the
final scaling results in PWIA and FSI might be different.

\section{Summary and Outlook}

In the present paper, we have expanded our modeling of
quark-hadron duality to describe a more realistic situation: we
now include the spin of the quark. Numerically, this is more
complicated than the all scalar model \cite{ijmvo} and the scalar
quark model \cite{jvod2}, where the solutions could be found
analytically. However, the inclusion of the quark spin paves the
way to study spin structure functions, which are one of the most
promising areas for the practical application of quark-hadron
duality.

We have tracked the changes introduced by making our models more
realistic. Most notably, adding spin to the quark does not seem to
influence the onset of scaling: our present model shows about the
same scaling behavior as the scalar quark model. This is in
contrast to the major changes in scaling behavior observed when
going from the all scalar model to the scalar quark model.

As before, we have seen that the features of duality observed in
experiments, namely scaling, oscillation of the resonances around
the scaling curve, and flat moments at higher momentum transfer,
are reproduced qualitatively. We have used a constituent quark
mass of $m = 258.46~MeV$ for our calculations. This number was
taken from a fit to heavy mesons \cite{waw}. However, nothing
hinges on using that particular value: we changed our quark mass
to $m = 10~MeV$, in order to have a value reminiscent of a current
quark mass, and repeated our calculations. It turns out that,
while scaling does set in a little faster, there are no
qualitative changes in the results.

In spite of the considerable numerical effort, we have not yet
reached full convergence in FSI. Therefore, the important question
if the FSI scaling curve coincides with the PWIA scaling curve
could not be answered with certainty yet. At the highest $q$ value
we reached, $q = 10~GeV$, we still see small differences between
FSI and PWIA curves. Further investigation of the high $q$ region
with different methods will be the subject of a future paper.

Our findings are different from the results in \cite{mpdirac}.
There, the author used massless quarks and a specific assumption
about the potential, namely an identical functional form for the
scalar potential $V_s$ and the vector potential $V_v$. The
condition $V_s = V_v$ leads to a simplification in the numerics,
as upper and lower components decouple. Results reported in
\cite{mpdirac} indicate a much larger difference between FSI and
PWIA results at $q = 10~GeV$. This may be due to the potential
chosen there.

For the first time, we have discussed several different
potentials. The combination of a scalar linear potential and a
vector color Coulomb potential (either with a running coupling
constant or without) can be considered a fairly realistic
approximation to nature. We were also able to demonstrate that the
observed qualitative features of duality- scaling, low $q$
duality, convergence of the moments - persist no matter which
potential is used. This hints at quark-hadron duality as a fairly
general property of inclusive electron scattering. While the shape
of the response functions is clearly influenced by the ground
state momentum distribution, and therefore by the potential, the
rate of convergence to scaling is unaffected by the choice of
potential, as demonstrated by studying the validity of the
y-scaling Callan-Gross relation.

Duality has been thoroughly explored experimentally in the
unpolarized nucleon structure functions, and new data on
polarization observables are available, too, both from Jefferson
Lab \cite{jlabspin} and Hermes at DESY \cite{hermes}. We will
apply our model to the calculation of spin structure functions
next.

{\bf Acknowledgments}: S.J. thanks the Theory Group of Jefferson
Lab for their kind hospitality. This work was supported in part by
funds provided by the U.S. Department of Energy (DOE) under
cooperative research agreement under No. DE-AC05-84ER40150 and by
the National Science Foundation under grant No. PHY-0139973.

\end{document}